\documentstyle[prd,aps,preprint]{revtex}
        		
\newcommand\ee{\end{equation}}
\newcommand\be{\begin{equation}}
\newcommand\eea{\end{eqnarray}}
\newcommand\bea{\begin{eqnarray}}

\newcommand\sunit{\,\mbox{s}}

\newcommand\km{\,\mbox{km}}

\newcommand\GeV{\,\mbox{GeV}}

\newcommand\Mpc{\,\mbox{Mpc}}

\newcommand\Mpl{M_{\rm Pl}}

\newcommand\lsim{\mathrel{\rlap{\lower4pt\hbox{\hskip1pt$\sim$}}
    \raise1pt\hbox{$<$}}}
\newcommand\gsim{\mathrel{\rlap{\lower4pt\hbox{\hskip1pt$\sim$}}
    \raise1pt\hbox{$>$}}}

\begin{document}

\preprint{LANCS-TH/9614, hep-ph/9609431}
\draft
\tighten

\title{Models of inflation and the spectral index \\
of the density perturbation}
\author{David H.  Lyth}
\address{School of Physics and Chemistry, \\
University of Lancaster, Lancaster LA1 4YB U.~K.}
\date{September 1996}
\maketitle

\begin{abstract}

If an adiabatic density perturbation is responsible for large scale 
structure and the cmb anisotropy, its 
spectral index $n$ will be measured in the forseeable future with an 
accuracy $\Delta n\sim .01$.
This is precisely the kind of accuracy required to 
distinguish between many models of inflation. 
Most of them have an inflationary potential
$V\simeq V_0(1\pm\mu\phi^p)$ with the constant term dominating.
Except for $0\lsim p\lsim 2$, the prediction  is 
$n=1\pm (2/N)(p-1)/(p-2)$, where $N$ is the number of $e$-folds of 
inflation after cosmological scales leave the horizon. It
typically lies in the range $0.9\lsim n\lsim 1.1$.
For $p=2$ one has $n=1\pm 2\Mpl^2 \mu$ where $\Mpl=(8\pi G)^{-1/2}$. 
A generic supergravity theory gives contributions of order
$\pm H^2$ to the inflaton mass-squared $m^2$ whereas inflation
requires $|m^2|\ll H^2$. Proposals for keeping $m^2$ small are surveyed
with emphasis on their implications for $n$. 
Finally, the case of a multi-component inflaton
(as in `double' inflation) is discussed.
In all cases, the best method of 
calculating the spectrum of the adiabatic density perturbation 
starts with the assumption that,
after smoothing on a super-horizon scale, the 
evolution of the universe along each comoving worldline 
will be practically the same
as in an unperturbed universe with the same initial inflaton field.
The possible isocurvature density perturbation is briefly discussed,
with a simple derivation of the fact that the
low multipoles of the cmb anisotropy are six times as big as for
an adiabatic density perturbation.

\end{abstract}

\section{Introduction}

Inflation generates an adiabatic density perturbation,
which is generally thought to be responsible
for large scale structure and the cosmic 
microwave background
(cmb) anisotropy, and it also generates gravitational waves which might
give a significant contribution to the latter
\cite{abook,kt,LL2}.

The spectrum of the density perturbation is conveniently 
specified by a quantity $\delta_H^2$ 
and the spectral index $n$ is defined by
$\delta_H^2\propto k^{n-1}$ where $k$ is the comoving wavenumber.
At present there is no detectable scale dependence, and 
observational limits on $n$
are only mildly constraining for inflationary 
models, the most notable result being that `extended' inflation 
\cite{extended}
is ruled out, except for rather contrived versions \cite{green}. 
But in the forseeable 
future one can expect a good measurement of $n$ 
and it is reasonable to ask 
what it will tell us.

At a purely phenomenological level the answer to the question 
is simple and well known. 
The spectral index $n$ is given by \cite{LL1,davis,salopek}
\be
n-1\simeq -3\Mpl^2(V'/V)^2 + 2\Mpl^2 (V''/V)
\label{n}
\ee
where $\Mpl=(8\pi G)^{-1/2}=2.4\times 10^{18}\GeV$ is the reduced Planck 
mass. The right hand side of this expression is a combination of the 
potential and its derivatives, 
evaluated at the epoch when
cosmological scales leave the horizon.
If $n$ is distinguishable from 1 the
combination will be measured, otherwise it will be constrained to near 
zero. Moreover, the relative contribution of 
gravitational waves to the mean-square low multipoles of the 
cmb anisotropy is given by \cite{rubakov,starob,d13,andrew}
$r\simeq 5\Mpl^2(V'/V)^2$. Their eventual detection or
non-detection will determine or constrain  the first term, so that measuring 
$n$ determines or constrains $V''/V$. Finally,
with more data one might measure the effective $n$ over a range of scales 
and also justify the use of more accurate formulas \cite{stly}, to obtain
limited additional information on the shape of $V$
while cosmological scales are leaving the horizon \cite{recon}.

This `reconstruction' approach is interesting, but it relies to a large 
extent on detecting gravitational waves and  will in any case always be 
limited by the narrow range of scales accessible to cosmology.
Meanwhile, 
since the advent of hybrid inflation \cite{l90,LIN2SC}, 
inflation model-building is beginning 
to come back into the fold of particle physics. Like the earliest models 
of inflation, but unlike those proposed in the intervening years,
hybrid inflation models work in the regime where all relevant fields
are small on the Planck scale \cite{CLLSW}. 
If one accepts the usual view that 
field theory beyond the standard model 
involves supergravity \cite{susy}, this represents a crucial simplification 
because in the context of supergravity the potential is expected to 
have an infinite power series 
expansion in the fields. For field values much bigger than $\Mpl$ 
one has no idea what form the potential will take, expect perhaps in the 
direction of the moduli fields of superstring theory
 \cite{bingall,paul}.
But in the small-field regime one is entitled to make the usual 
assumption, that the expansion is dominated by a very few low-order
terms. As a result, inflation model-building has  become a more varied,
yet better controlled activity.

Given this situation it is reasonable to go beyond the purely 
phenomenological level, and ask how well a measurement of the 
spectral index will discriminate between different models of inflation.
That
question is the main focus of the present paper.

There are of course many other aspects of inflation model-building.
The prediction Eq.~(\ref{delh})
of the normalization of $\delta_H$  provides another constraint 
on the parameters of the inflationary potential, which can crucially 
effect the viability of a given potential. It may be regarded
as fixing the magnitude of $V$ as opposed to  its derivatives,
a quantity whose theoretical significance is not yet clear
(for some recent discussions 
see Refs.~\cite{CLLSW,mutated,paul,lisa,graham}).
Another theoretical issue is the
difficulty of implementing slow-roll inflation in the context of 
supergravity \cite{CLLSW,ewansgrav}.
These two things will be considered
briefly in the present paper, but other
aspects of inflation model building, in particular reheating 
and `preheating' (see for instance Ref.~\cite{kls}) 
will not be discussed at all.

In most of the discussion 
we assume that the slowly-rolling inflaton field is essentially 
unique. This is the case in most of the models so far proposed,
and is necessary for the validity of the above formula for $n$.
But in Section VII we discuss the case
where there is a family of inflationary trajectories lying
in the space of two or more fields; in other words, where there is a 
multi-component inflaton. The quantum fluctuation kicks the 
field from one trajectory to another, and if
the trajectories are physically 
inequivalent this gives a contribution 
to the spectrum of the density perturbation which has to be added to the 
usual one, coming from the fluctuation back and forth along the same 
trajectory. In all cases, the best method of 
calculating the spectrum of the adiabatic density perturbation 
starts with the assumption that,
after smoothing on a super-horizon scale, the 
evolution of the universe along each comoving worldline 
will be the same
as in an unperturbed universe with the same initial inflaton field.
We also discuss briefly the 
case of an isocurvature density perturbation,
giving a simple derivation of the fact that the
low multipoles of the cmb anisotropy are six times as big as for
an adiabatic density perturbation.

We assume that the scale dependence of 
$\delta_H$ can be represented by a spectral index, $\delta_H^2\propto
k^{n-1}$, at least while cosmological scales are leaving the horizon.
This is the case if slow-roll inflation is continuously 
maintained, unless there is a cancellation between the terms of 
Eq.~(\ref{n}) for $n-1$. A feature, such as a shoulder in the spectrum,
can easily be generated if slow-roll inflation is briefly interrupted 
at the epoch when the relevant scale is leaving 
the horizon \cite{interrupt,c3,d7,d9,d10,d13}.
However to get any kind of feature during the few $e$-folds 
corresponding to cosmological scales requires a delicate balance of 
parameters. Observation does suggest a shoulder in the spectrum of the 
{\em present} density perturbation, in the 
tens of $\Mpc$ range, though at present the evidence is also consistent
with a smooth rise corresponding to $n\simeq 0.8$.
But the spectrum $\delta_H$ refers to the density perturbation in
the early universe, and the  most natural explanation for a shoulder 
in the spectrum of the observed density perturbation would be 
that it is in the transfer function relating the present density 
perturbation to the early one.
Indeed, any departure from the pure cold dark 
matter hypothesis leads to a shoulder on just the required scales.
The conclusion is that there is little motivation to 
depart from the slow-roll inflation paradigm, and its attendant
scale dependence $\delta_H^2\propto k^{n-1}$.

The discussion is confined to models of inflation leading to a spatially
flat universe on observable scales. This is the generic case,
though at the expense of moderate fine-tuning the 
universe can be open (negatively curved) if we live inside a bubble formed at 
the end of a preliminary era of 
inflation \cite{open}.

Throughout the paper, $a$ denotes the scale factor of the universe
normalized to 1 at present and $H=\dot a/a$ is the Hubble parameter,
with present value $H_0=100h\km\sunit^{-1}\Mpc^{-1}$. The corresponding 
Hubble distance is $cH_0^{-1}=3000h^{-1}\Mpc$.
From now on the units are $c=\hbar=1$.
The comoving wavenumber in a Fourier series expansion is denoted
by $k/a$, so that $k$ is the present wavenumber and $k^{-1}$
conveniently specifies the corresponding comoving scale.
The scale is said to be outside the horizon when $aH/k$
is bigger than 1 and inside it otherwise. A given scale of interest 
leaves the horizon during inflation and re-enters it long afterwards
during radiation or matter domination.

The plan of the paper is as follows. In Section II the eventual accuracy 
of the measurement of $n$ is discussed. In Section III some general 
issues relating to inflation model-building are discussed, and 
specific models are discussed in Section IV and V. In Section VI
various ways of implementing slow-roll inflation in the context of 
supergravity are briefly discussed, focussing on their implications for the 
spectral index. Section VII discusses inflation with a multi-component
inflaton, and the calculation of the spectrum of the density 
perturbation in both this and the single-component case. A
summary is provided in Section VIII. 

\section{How accurately will the spectral index be determined?}

Before discussing the theoretical predictions we need an idea of the 
eventual accuracy to expect for the determination of the 
spectral index of the adiabatic density perturbation. First let us define 
this perturbation.

\subsection{The spectrum of the adiabatic density perturbation}

The evolution of small density perturbations in the universe is 
governed by linear equations, with each Fourier mode
evolving independently. On scales well outside the horizon
the evolution is governed purely by gravity, and in each comoving
region its evolution with respect to the local proper time
is is the same as it would be in an unperturbed universe.

For the moment we are not interested in the
evolution during inflation and in the very early universe. Rather,
we focus on the relatively late era when 
the non-baryonic dark matter has its present form, and the
radiation-dominated era immediately preceding the present
matter-dominated era has begun.
During this era, consider the perturbations $\delta\rho_i$ in the energy 
densities of the various particle species. They are said to be adiabatic 
if each of the densities $\rho_i$ is a unique function 
of the total density $\rho$. The perturbations all vanish if we slice 
the universe into 
hypersurfaces of constant energy density. With any other slicing,
all species of matter have a common density contrast and so do all 
species of radiation, with $\delta\rho_m/\rho_m=\frac34\delta\rho_r
/\rho_r$. 

A different type of density perturbation would be an isocurvature one, 
for which the total density perturbation vanishes but the separate ones
do not. The vacuum fluctuation during inflation  will generate an adiabatic 
perturbation, which is the subject of the present paper. At the end of 
Section VII we consider briefly the isocurvature perturbation that might 
also be produced.

The 
adiabatic density perturbation is determined by the total density 
perturbation $\delta\rho/\rho$. This in turn is determined
by a 
quantity $\cal R$,
which defines the curvature of hypersurface orthogonal to 
comoving worldlines (`comoving hypersurfaces').\footnote
{For each Fourier mode, the
curvature scalar $R^{(3)}$ is given by 
${\cal R}= \frac14 (a/k)^2 
R^{(3)}$ (it is a perturbation since the unperturbed universe is 
supposed to be flat).
There is unfortunately no standard notation for $\cal R$.
It was first defined in Ref.~\cite{bardeen80}
where it was called $\phi_m$. It was called ${\cal R}_m$ in 
Ref.~\cite{kodamasasaki}, and is a factor $\frac32k^{-2}$
times the quantity $\delta K$ of Ref.~\cite{lyth85}.
On the scales far outside the horizon where it is constant
(the only regime where it is of interest)
it coincides with the quantity $\xi/3$ of Ref.~\cite{bst}
and the quantity $\xi$ of Ref.~\cite{c3}.}
At least during the relatively late era that we are focussing on at the 
moment, $\cal R$ is time-independent, and 
\be
\frac{\delta\rho}{\rho}=\frac49 \left(\frac{k}{aH}\right)^2 {\cal R}
\ee
(As always we focus on a single Fourier component, well before horizon 
entry.)

After horizon entry the evolution becomes more complicated, and
for radiation as opposed to matter it 
involves the full phase-space densities. The main  observational quantities
are
the present matter density perturbation,
which we can get a handle by observing the distribution and motion of
galaxies, and the cmb anisotropy. There is also some 
information on the matter density 
perturbation at earlier times.

Inflation typically predicts
that $\cal R$ is a a Gaussian random field, which is usually assumed and 
is consistent with observation. Then all stochastic properties
of the perturbations 
are determined by the spectrum ${\cal P}_{\cal R}$ of $\cal R$.\footnote
{The spectrum ${\cal P}_f(k)$ 
of a quantity $f$ is the modulus-squared of its Fourier
coefficient, times a factor which we choose so that 
$f$ has mean-square value $\bar{f^2}=\int^\infty_0 {\cal P}_f
dk/k$.}
We are supposing that any scale dependence of the spectrum
has the form ${\cal P}_{\cal R}\propto k^{n-1}$ where $n$ is the 
spectral index.

Instead of ${\cal P}_{\cal R}$ one usually considers
\be
\delta_H^2\equiv (4/25) {\cal P}_{\cal R}
\label{delhdef}
\ee
Then, on scales 
$k^{-1}\gsim 10\Mpc$ where its evolution is still linear, the spectrum
of the present density perturbation $\delta\equiv\delta\rho/\rho$ 
is given by 
\be
{\cal P}_\delta = \delta_H^2 (k/H_0)^4 T^2(k)
\ee
The transfer function $T(k)$ represents non-gravitational effects,
and is equal to 1 on the large scales 
$k^{-1}\gg 100\Mpc$ where such effects are absent.
Information on ${\cal P}_\delta$ 
on scales of order $10$ to $100\Mpc$ is provided
by a variety of observations on the abundance, distribution and
motion of galaxies, and if the transfer function is known
one can deduce $\delta_H$.

The $l$th multipole of the cmb anisotropy 
probes the scale 
$k^{-1}\simeq 2/(H_0 l)= 6000h^{-1}l^{-1}\Mpc$, corresponding to the 
distance subtended by angle $1/l$ at the surface of last scattering.
Since this `surface' is of order $10\Mpc$ thick the anisotropy will be 
wiped out on smaller scales, corresponding to $l\gsim 10^4$.
It thus lies entirely within the linear regime.
The 
mean-square multipoles seen by a randomly placed observer
can be calculated in terms of
$\delta_H^2(k)$ at $k^{-1}\simeq 2/(H_0 l)$, given another 
transfer function encoding the effect of non-gravitational
interactions. At present
observations exist only for $l\lsim 100$, corresponding to
$k^{-1}\gsim 100\Mpc$.

The COBE measurement corresponds to 
multipoles $l\sim 30$, and gives an accurate determination of 
$\delta_H$ on the corresponding scales because the evolution is
purely gravitational (dominated by the Sachs-Wolfe effect).
On the
scale $k\simeq 5H_0$ corresponding to the
$l\simeq 10$ multipoles which define the centre of the measured
range one finds
   \cite{delh,bunnwhite}
\be
\delta_H= 1.94\times 10^{-5}
\label{cobe}
\ee
with a $2\sigma$  uncertainty of $15\%$.
This assumes that gravitational waves give a negligible contribution.
Otherwise, the result is multiplied by a factor
$(1+r)^{-1/2}$ which can be calculated in a given model of 
inflation.

In order to arrive at an observational value for
$n$ 
we need measurements of $\delta_H$ also on smaller scales. Here
transfer functions are needed and they 
depend strongly on
the cosmology.
Since we are taking particle physics
seriously it is reasonable to assume that the cosmological constant
vanishes and (in the interests of simplicity) that the density is
critical. Then 
 the main uncertainties
are the value of the Hubble constant 
$H_0=100h\km\sunit^{-1}\Mpc^{-1}$,
the value of the 
baryon density $\Omega_B$, and the 
nature of the dark matter. 

The simplest assumption is pure cold dark matter. This case is viable at 
present \cite{constraint2} with $h\simeq 0.5$, $\Omega_B\simeq 0.12$ and
\be
0.7 \lsim n \lsim 0.9
\ee
This assumption will be invalidated if a shoulder is observed 
in the spectrum, or if
$h$ or $\Omega_B$ differ from the above values by more than $10\%$
or so.

A more robust 
assumption is that there is an admixture
of hot dark matter in the form of a single neutrino species
\cite{mdm},  and taking
$\Omega_B<.15$ and $h>0.4$ 
a result at something like 95\% confidence-level is
\cite{constraint1,constraint2}
\be
0.7< n < 1.3
\label{nbound}
\ee
The uncertainty in $n$ comes mostly from the uncertainties in 
$H_0$, $\Omega_B$ and the nature of the dark matter.
The quoted lower bound comes from the cmb anisotropy in the region of the 
first Doppler peak and depends only on $\Omega_B$, though other 
data 
point to a lower bound that is not much weaker.
The upper bound depends strongly on $h$ and the nature of the dark 
matter. With the same hypothesis about the dark matter one finds
$n<1.04$ if $h>0.5$ and $n<0.9$ if  $h>0.6$, a very dramatic 
improvement. The bounds on $n$ under other hypotheses about the dark 
matter have not been explored in detail but one does not expect wildly 
different results since the effect of modifying pure cold dark 
matter is always to reduce the density perturbation on small scales.

The bounds on $n$ become tighter if there are significant gravitational 
waves, but that is not the case in small-field models, nor is it the 
case in most other models that have been proposed \cite{mygwave}.
In models where $r\simeq 5(1-n)$
(corresponding to an exponential potential, and approximately to
a power-law potential) the lower bound becomes $n>0.8$
\cite{constraint1}, which 
rules 
out extended inflation \cite{extended} except for rather 
contrived versions \cite{green}.

To get an idea of the eventual accuracy to be expected for $n$, suppose that 
$\delta_H$ has been measured at just
two wavenumbers $k$ and $q$, with $k>q$. 
 If the values of $\delta_H$ have fractional uncertainties
$\Delta_k$ and $\Delta_q$, the uncertainty in $n$ is \cite{knox}
\be
\Delta n = (\Delta_k^2 +\Delta_q^2)^{1/2} / \ln(k/q)
\ee
Other things being equal one will use the longest possible 
`lever arm' $k/q$, so the idealization of using only two wavenumbers
is not too far from reality.

The longest useful lever arm at present 
is provided by $k= (10h^{-1}\Mpc)^{-1}$, the smallest scale that
is linear now, and $q= 5H_0= (600h^{-1}\Mpc)^{-1}$,
the scale effectively probed by COBE.
If each piece of data has 
a 10\%  uncertainty this gives $\Delta n =.034$.
At present the uncertainties are $\Delta_q= 15\%$ as quoted above, and 
{\em excluding the uncertainty in the transfer function} 
$\Delta_k=30\%$ \cite{viana}. Thus we would already be  not too far
from the situation of $10\%$ errors if only the cosmological parameters 
and the nature of the dark matter were understood.
It seems quite reasonable to suppose
that good progress in this direction will have been made in a few years,
giving an accuracy of order 10\% on both pieces of data and an accuracy
$\Delta n\simeq .04$ for the spectral index.

Looking for much higher accuracy, one cannot use
data corresponding to multipoles of order $10$. The reason is that
the value of $\delta_H$ adduced even from perfect data is subject
to cosmic variance, arising from the fact that the multipoles are
measured only at our position. Using multipoles of order $l$ 
(over unit interval of $\ln l$) the uncertainty due to cosmic variance
is of order $1/l$. To achieve something like 1\% accuracy, one needs
data centered on  $l= 100$, corresponding 
to
$q=(60h^{-1}\Mpc)^{-1}$.
With  1\% uncertainty also at $k=(10h^{-1}\Mpc)^{-1}$,
 one gets  $\Delta n
=.008$. We see that  improving the accuracy by a factor 
10 only increases the accuracy on $\Delta n$ by a factor 4, because
of the necessarily shorter lever arm. 

In the future one will be able to go to smaller scales, with
the best determination of the spectral index coming  from
the cmb anisotropy with good angular resolution, but 
an accuracy better than $\Delta n\sim .01$
will be difficult to achieve if only because of the uncertainty in the 
cosmological parameters \cite{knox,jungman}. 
Fortunately, this accuracy turns out to be all that is needed
to discriminate sharply between different models.

\subsection{Scale-dependence of the spectral index?}

If the power-law parametrization $\delta_H^2\propto k^{n-1}$ is 
imperfect one can still define an effective spectral index by
$n-1\equiv d\ln\delta_H/d\ln k$ and it then makes sense to 
ask whether the variation of the spectral index will ever be 
observable. 

In simple models 
$n-1$ is proportional to $1/N$ if it varies at all, 
where $N$ is the number of $e$-folds of 
inflation remaining when the relevant scale leaves the horizon,
and we consider only that case. 

Although the smallest scale on which $\delta_H(k)$ can be measured is
at present about $10\Mpc$, this limit will go down an order of magnitude 
or two in the future, making the total
accessible range perhaps four decades ranging from 
$6000h^{-1}\Mpc$ to say $0.6h^{-1}\Mpc$. This would
 correspond to $\Delta \ln k =4.5$ (measured from the centre of the 
range) and therefore to $\Delta N=4.5$.

As we shall discuss in the next section, the number of $e$-folds
after (say) the scale in the centre of the this range leaves the horizon 
depends on the history of the universe after inflation, as well as
the energy scale at which inflation ends. For a
high (ordinary) inflationary 
energy scale $N\simeq 50$ is reasonable, but a low scale combined with 
an
era of thermal inflation \cite{thermal}
will give $N\simeq 25$. One or more 
additional bouts of thermal inflation are not unlikely 
from a theoretical viewpoint, but since each one reduces $N$ by
$\simeq 10$ there had better be at most one or two.

Assume first at most one bout of thermal inflation. Then the 
fractional 
changes in $n$ and $N$ are small so that 
\be
\frac{\Delta n}{|1-n|} = \frac{\Delta N}{N}
\ee
Let us suppose that the range is split in half to yield two separate
determinations of $n$, corresponding to $\Delta N=4.5$.
Then one finds that 
\be
\frac{\Delta n}{.01} =\frac{|1-n|}{0.11}\frac{N}{50} 
\ee
From what has been said above, the scale dependence 
$\Delta n$ will have to be bigger
than $.01$ if it is to be detectable in the forseeable future.
As a result its variation will be detectable only if it is quite far 
away from 1.

Now suppose that there are two bouts of thermal 
inflation so that $N$ varies from say 10 to 20 while cosmological scales 
leave the horizon. Then $n-1$ will double over the observable range
so that its variation might be detectable even if $|n-1|$ is of order
$.01$. Three bouts would of course have a dramatic effect, with
four probably forbidden because cosmological scales would re-enter the 
horizon during thermal inflation which would lose the prediction for the
density perturbation.

If the spectral index is strongly scale-dependent it is more useful
to consider the spectrum itself. 
The case that we are considering, $n-1=q/N$, corresponds to
$\delta_H^2\propto N^q$. As discussed in Section VIII, the 
strongest scale dependence obtained from a reasonably motivated
potential corresponds to $q=-4$, coming from a cubic potential. 
In order to produce enough early 
structure, $\delta_H$ should fall by no more than a factor 5 or so as 
one goes down through the four orders of magnitude available to 
observation \cite{constraint1}, which with the cubic potential 
corresponds to requiring $N\gsim 11$ in the center of the range.
Thus the cubic potential 
permits at most one additional bout of thermal inflation,
at least with a low scale for ordinary inflation.
Even if $N$ is not low enough to make it dramatic, the scale-dependence
of $n$ in this model might still be enough to improve the 
viability of a pure cold dark matter model \cite{grahamnew}.

\subsection{Observational constraints on very small scales}

The above considerations apply only on cosmological scales. On smaller scales 
there are only upper limits on $\delta_H$. The most useful 
limit, from the viewpoint of constraining models of inflation,
is the one on the smallest relevant scale which is the one leaving the
horizon just before the end of inflation. It has been considered
in Refs.~\cite{jim} and \cite{lisa}, and according to the latter reference it 
corresponds to $\delta_H\lsim 0.1$. 
In models where $n$ is constant this corresponds
 (using Eq.~(\ref{cobe}))
to $n-1<0.33(50/N)$, which is no stronger thant the constraint 
coming from cosmological scales.
It represent a powerful additional constraint in models where
$n$ increases on small scales, as discussed in an extreme case in
Ref.~\cite{lisa,glw}.

\section{Slow-roll inflation}

Before discussing particular models 
it will be useful to deal with some generalities.
Practically all models are of the 
slow-roll variety \cite{kt,abook,LL2}. Such models are the
simplest, and at least while cosmological scales are leaving the horizon 
they are motivated by the observed fact that the spectrum has mild scale
dependence, which is consistent with a power law.
Until Section VII we focus
on models in which the slow-rolling 
inflaton field $\phi$ is essentially unique.

During inflation the
potential $V(\phi)$ is supposed to 
satisfy the flatness conditions $\epsilon\ll 1$ and $|\eta|\ll1$, 
where
\bea
\epsilon&\equiv &\frac12\Mpl^2(V'/V)^2\\
\eta&\equiv &\Mpl^2 V''/V
\eea
Given these conditions, the evolution 
\be
\ddot\phi+3H\dot\phi=-V'
\label{phiddot}
\ee
typically settles down to the slow-roll evolution
\be
3H\dot\phi=-V'
\label{slowroll}
\ee

Slow roll and the flatness condition $\epsilon\ll 1$
 ensure that the energy density 
$\rho=V+\frac12\dot\phi^2$ is close to $V$ and is slowly varying.
As a result $H$ is slowly varying,
which implies that one can write
$a\propto e^{Ht}$ at least over a Hubble time or so.\footnote
{Here and in what follows,
slow variation of a function
$f$ of time means $|d\ln f/d\ln a|
\ll 1$, and slow variation of a function $g$ of wavenumber means
that $|d\ln g/d\ln k|\ll 1$.}
The
flatness condition $|\eta|\ll 1$ then ensures that 
$\dot\phi$ and $\epsilon$ are slowly
varying.

A crucial role is played by the number of Hubble times
$N(\phi)$ of inflation, still remaining when $\phi$ has a given value.
By definition $dN=-H\,dt$, and 
the 
slow-roll condition together with the flatness condition
$\epsilon\ll 1$ lead to 
\be
\frac{dN}{d\phi}=-\frac{H}{\dot\phi}=\frac{V}{\Mpl^2V'}
\label{Nrelation}
\ee
This leads to 
\be
N=\left| \int^\phi_{\phi_{\rm end}} \Mpl^{-2}\frac V{V'} d\phi
\right|
\label{nint}
\ee

The slow-roll paradigm is motivated, while cosmological 
scales are leaving the horizon, by the observed fact that 
$\delta_H$ does not vary much on such scales. In typical models
slow-roll persists until almost the end of inflation, but it can 
happen that it fails much earlier. The loop-correction model of
Ref.~\cite{ewanloop} is the only one so far exhibited with this property,
and as discussed there the only effect of the failure of slow roll is
that one needs to replace the above expression for $N(\phi)$ with 
something more accurate, when it is used to work out the value of
$\phi$ corresponding to cosmological scales.\footnote
{In principle one might be 
concerned that the quantum fluctuation during non-slow-roll inflation
might become big and produce unwanted black holes. But in the 
slow-roll regime the spectrum $\delta_H$ becomes {\em smaller }
as the speed of rolling is increased (ie., as $\epsilon$ is increased)
and it seems reasonable to suppose that this trend persists beyond the 
slow-roll regime.}

\subsection{The predictions}

The quantum fluctuation of the inflaton field
gives rise to an adiabatic density
perturbation, whose spectrum is
\be
\delta_H^2(k) = \frac1{75\pi^2\Mpl^6}\frac{V^3}{V'^2}
=
\frac1{150\pi^2\Mpl^4}\frac{V}{\epsilon}
\label{delh}
\ee
In this expression, the
potential and its derivative are evaluated at the epoch of 
horizon exit for the scale $k$, which is defined by $k=aH$.

This result was in essence derived at about the same time in 
Refs.~\cite{hawking,starob82,guthpi,bst}.
To be precise, these authors give results which lead more or less
directly to the desired one after the 
spectrum has been defined, though that last step is not 
explicitly made and except for the last work only a 
particular potential is discussed. 
(The last three works give results equivalent 
to the one we quote, and 
the first gives a result which is approximately the same.)
Soon afterwards it was derived again, this time with an explicitly
defined spectrum \cite{lyth85}.\footnote
{Strictly 
speaking none of these five derivations is completely satisfactory.
The first three make simplifying assumptions, 
and all except the third assume
something equivalent to the constancy of $\cal R$ 
without adequate proof. But as discussed in 
Section VII the constancy of $\cal R$, and therefore the validity of the result,
is easy to establish 
for a single-component inflaton.
Also, none of them properly considers the effect of the inflaton field
perturbation on the metric, but again this is easily justified to
first order in the slow-roll approximation \cite{LL2}. 
A little later a formalism was given that takes the metric perturbation into
account \cite{mukhanov,sasaki}, which formed the basis for the more
accurate calculation of \cite{stly}.}

Inflation also generates
gravitational waves, whose relative contribution to the mean-squared
low multipoles of the cmb anisotropy is
\be
r = 5\Mpl^2(V'/V)^2
\ee
The contribution of the gravitational waves was estimated
roughly in Ref.~\cite{rubakov}, and in 
Ref.~\cite{starob} a result equivalent to $r=6.2\Mpl^2(V'/V)^2$
was obtained under certain approximations. The more correct factor
$\simeq 5$, for the multipoles most relevant for the determination of 
the COBE normalization of $\delta_H$, was given in Refs.~\cite{d13,andrew}.

Returning to the spectrum of the density perturbation, comparison
of the prediction
(\ref{delh}) with the value deduced from the COBE observation of the cmb 
anisotropy gives (\ref{cobe}), ignoring the gravitational waves\footnote
{In this context 
there is no point in 
including the effect of gravitational waves on the cmb anisotropy,
since
the prediction for $\delta_H$ that is being used has an error of at least
the same order. If necessary one could include the effect 
of the gravitational waves using the more accurate formula for
$\delta_H$ \cite{stly,recon}.}
\be
\Mpl^{-3} V^{3/2}/V' = 5.3\times 10^{-4}
\label{cobecons}
\ee
This relation provides a useful constraint
on the parameters of the potential. It can be written in the equivalent 
form
\be
V^{1/4}/\epsilon^{1/4}=.027\Mpl=6.7\times 10^{16}\GeV
\label{vbound}
\ee
Since $\epsilon$ is much less than 1, the 
inflationary energy scale $V^{1/4}$
is at least a couple of orders of magnitude
below the Planck scale \cite{lyth85}.

Our main focus is on the 
spectral index of the density perturbation. It may be defined by
$n-1\equiv 2d\ln \delta_H/d\ln k$. 
Since the right hand side of
Eq.~(\ref{delh}) is evaluated when $k=aH$, and since
the rate of change of 
$H$ is negligible compared with that of $a$ we have
$d\ln k=H dt$. From the slow-roll condition
$dt=d\phi/\dot\phi=-(3H/V')d\phi$
and therefore
\be
\frac{d}{d\ln k}
=-\Mpl^2 \frac{V'}{V} \frac{d}{d\phi}
\label{nexpression}
\ee 
This gives
\cite{LL1,davis,salopek} the formula 
advertized in the introduction,
\be
n-1=-6\epsilon+2\eta
\label{n2}
\ee

More accurate versions of the formulas for $\delta_H$ and $n$ have 
been derived \cite{stly} which provide estimates of
the errors in the usual formulas. They are
\bea
\frac{\Delta\delta_H}{\delta_H}&=&O(\epsilon,\eta)\\
\Delta n &=& O(\epsilon^2,\eta^2,\gamma^2)
\label{deltan2}
\eea
where $\gamma$ is a third flatness  parameter defined by
$\gamma^2\equiv \Mpl^4 V'''V'/V^2$.
We see that the usual formula $n-1=-6\epsilon+2\eta$ is valid
(barring cancellations) if and only if 
there is a third flatness condition
(in addition to $\epsilon\ll 1$ and $|\eta|\ll 1$) which is
\be
\gamma^2\ll\max\{\epsilon,\eta\}
\label{flat3}
\ee
In typical models it
is an automatically consequence of the other two,
but it is in principle an additional requirement.

In most models, $\epsilon$ and $\gamma$ are 
both negligible compared with $\eta$. Then 
$n=1+2\eta$, and
\be
\Delta n\sim |1-n|^2
\ee
This theoretical error 
will be comparable with the eventual observational uncertainty 
(say $\Delta n\sim .01$) if 
it turns out that $|1-n|\gsim 0.1$, in which case the use of the
more accurate formula would be needed to fully interpret the 
observation. However, as will become clear an accuracy
$\Delta n$ of order $.01$ is (in the models proposed so far)
needed only if $|1-n|\lsim 0.1$. Thus the 
more accurate formula for the spectral index is not required for the 
purpose of distinguishing between models.

Differentiating the usual expression one finds \cite{running}
\be
\frac{d\ln (n-1)}{d\ln k}= [-16\epsilon\eta
+24\epsilon^2 + 2\gamma^2]/(n-1)
\label{running}
\ee
Barring cancellations, we see that 
$n-1$ varies little over an interval
$\Delta \ln k\sim 1$ if and only if all three flatness conditions are 
satisfied.
Cosmological scales correspond to only $\Delta \ln k\sim 10$, 
and in most models $n-1$ varies little over this interval too. Then
one can write $\delta_H^2\propto k^{n-1}$, the usual
definition of $n$. (A model where the two terms of $n-1$ cancel,
allowing it to vary significantly, is mentioned in Section V(D).)

\subsection{A simpler formula for the spectral index}

In almost all models proposed so far,
$\epsilon$ is negligible so that one can write
$n-1=2\eta$.

Consider first the potential
$V\simeq V_0(1\pm\mu\phi^p)$,  
with the constant term dominating. We shall encounter potentials of this 
form frequently, and since the constant term dominates they give
\be
\frac{\epsilon}{\eta}=\frac{p}{p-1}\frac\mu2 \phi^p\ll \frac p{p-1}
\ee
Except in the case $p\simeq 1$, we see that $\epsilon$
is indeed negligible in these models.

This result applies whether or not $\phi$ is small on the Planck scale. 
If it is small, one 
can argue that $\epsilon$ is negligible irrespectively of the form of 
the potential.
To see this, use
the estimate $\Delta \ln k\simeq 9$ for the range of cosmological 
scales. It corresponds to $9$ $e$-folds of 
inflation.
 In slow-roll inflation the quantity $V'/V$ has negligible variation over one 
$e$-fold and in typical models it has only small variation over the
9 $e$-folds. Taking that to be the case, one learns from 
(\ref{Nrelation})
that in small-field models the contribution of $(V'/V)$ to the spectral 
index is $3\Mpl^2 (V'/V)^2\ll (1/9)^2=.04$,
which is indeed negligible.

This upper bound on $(V'/V)$ means \cite{mygwave}
that the effect of gravitational 
waves on the cmb anisotropy will be too small ever to detect,
because owing to cosmic variance the relative 
contribution
$r\simeq 5\Mpl^2(V'/V)^2$ 
has to be $\gsim 0.1$ if it is to be detectable 
\cite{turner}.
The effect is actually undetectable
in most large-field models as well \cite{mygwave}.

\subsection{The number of $e$-folds}

A model of inflation will give us an inflationary potential $V(\phi)$,
and a prescription for the value $\phi_{\rm end}$ of the field at
the end of inflation. What we need to obtain  the predictions
is the value of $\phi$ when cosmological scales 
leave the horizon, which is given in terms of $N$ by
Eq.~(\ref{nint}). 
This quantity can in turn be 
determined if the history of the universe after inflation 
is known.
Consider first the epoch when the 
scale
$k^{-1}=H_0 ^{-1}= 3000h^{-1}\Mpc$ leaves the horizon, which can be 
taken to mark the 
beginning of cosmological inflation.\footnote
{This is  roughly the biggest cosmologically interesting
scale. The absolute limit of direct observation is $2H_0^{-1}$,
the distance to the particle horizon. Since
 the prediction is made for a randomly placed observer
 in a much bigger  patch,
 bigger scales in principle contribute to it,
but sensitivity rapidly decreases outside our horizon.
Only if the spectrum increases sharply on very large scales
\cite{gz} might there be a significant effect.}
Using a subscript 1 to denote this epoch, $N_1= \ln(a_{\rm end}/a_1)$.
Since this scale is the one  entering the horizon now,
 $a_1H_1=a_0H_0$ where the subscript 0 indicates the present
epoch and 
\be
N_1=
\ln\left(\frac{a_{\rm end}H_{\rm end}}{a_0H_0}\right)
-\ln\left(\frac{H_{\rm end}}{H_1}\right)
\ee
The second  term  will be given by the model of inflation and is usually
 $\lsim 1$; for simplicity let us 
ignore it. The first term  is known if we
know the evolution of the scale factor between the end of cosmological
inflation and the present. 
Assume first that the end of inflation gives way promptly to
matter domination, which is followed by a radiation dominated era
lasting until the present matter dominated era begins,
one has \cite{LL2}
\be 
N_1=62-\ln(10^{16}\GeV/V_{\rm end}^{1/4}) 
-\frac13\ln(V_{\rm end}^{1/4}/\rho_{\rm reh}^{1/4})
\ee
($\rho_{\rm reh}$ is the `reheat' temperature, when radiation domination 
begins.)
With $V^{1/4}\sim 10^{16}\GeV$ and instant reheating this gives
$N_1\simeq 62$, the biggest possible value. However,
$\rho_{\rm reh}$ should probably be no bigger than
$10^{10}\GeV$ to avoid too many gravitinos, and using that value
gives $N_1=56$, perhaps the biggest reasonable value.
With $V^{1/4}=10^{10}\GeV$, the
lowest scale usually considered, one finds $N_1=47$ with instant reheating,
and $N_1=39$ if reheating is delayed to just before nucleosynthesis. 
If there is in 
addition an era of thermal inflation 
\cite{thermal} lasting about 10 $e$-folds this figure is reduced to
$N_1\simeq 29$. Subsequent eras of the thermal inflation are 
not particularly unlikely from a theoretical viewpoint,
but since each era would subtract $\simeq 10$ from $N_1$
this would have dramatic observational consequences with
more than two eras practically forbidden. 

The smallest scale that will be directly probed in the forseeable
future is perhaps 
four orders of of magnitude lower than $H_0^{-1}$, which corresponds to
 replacing $N_1$ by $N_1-9$. 
A central value, appropriate for  use when  calculating 
the spectral index, is therefore $N\simeq N_1-4.5$,
and assuming at most one bout of thermal inflation
it probably lies in the range
\be
24\lsim N \lsim 51
\ee

In the models that have been proposed, the predicted value of $n-1$
is either independent of $N$, or is proportional to $1/N$.
In the latter case, taking $N=25$ instead of the usual $N=50$
or so doubles the predicted value of $n-1$.

\subsection{Hybrid and non-hybrid models of inflation}

In non-hybrid models of inflation, the potential $V(\phi)$
is dominated by the inflaton field $\phi$. The potential
has a minimum corresponding to the vacuum value (vev) of $\phi$,
at which $V$ vanishes (or anyhow is much 
smaller than during inflation).
Inflation ends when the minimum is approached, leading to a
failure of one of the flatness conditions, and 
$\phi$ then oscillates about its vev.
In hybrid 
models, the potential $V(\phi,\psi)$
is dominated by some other field $\psi$,
which is held in place by its interaction with $\phi$. 
In ordinary hybrid models $\psi$ is fixed, whereas in 
`mutated' models it is slowly varying as it adjusts to minimize the 
potential at fixed $\phi$, but either way we again
end up with an effective 
potential $V(\phi)$ during inflation. However, 
inflation now typically ends 
when the other field is destabilized, as 
$\phi$ passes through some critical value $\phi_c$, rather than by
failure of the flatness conditions. After inflation ends
both $\phi$ and the other field oscillate about their vevs, 
though they might be very inhomogeneous depending on the 
model.

\subsection{The power-series expansion}

In the context of supergravity one expects the potential $V$ to have
an infinite power series expansion in the fields \cite{susy}. 
It vanishes 
at its minimum, which corresponds to the vacuum expectation values
(vevs) of the fields. 
Assuming for simplicity
that odd terms are forbidden by some symmetry, the expansion for a 
single field with all others fixed at their vev 
will be of the form (with
coefficients of either sign)
\be
V(\phi)
=V_0 +\frac12 m^2\phi^2 +\frac14\lambda\phi^4+\lambda' \Mpl^{-2}\phi^6 
+\cdots
\label{power}
\ee
The minimum, at which $V$ vanishes, corresponds to the vev of
$\phi$. 

In the usual applications of particle theory,
to such things as the standard model, neutrino masses, Peccei-Quinn 
symmetry or a GUT, one is interested only in field values
$\phi\ll \Mpl$. In this regime it is reasonable to expect
the series to be dominated by one, two or at most three terms.
To be precise, one expects the quadratic and/or the quartic
terms to be important, 
plus at most one higher term. As a result there is good control 
over the form of the potential, and plenty of predictive power.
It is attractive to suppose that inflation model-building lies in the 
same regime, and indeed one may even hope that the fields 
occurring in the inflation model are ones that already occur in some 
particle physics application.\footnote
{For a recent proposal for identifying the inflaton field with a flat 
direction of a supersymmetric extension of the standard model,
see Ref.~\cite{lisa}; note though that the
non-inflaton field is a distance of order $\Mpl$ from its vacuum value
in the models considered there.}

As one considers bigger field values the justification for keeping
only a few terms becomes weaker. For values of order $\Mpl$ one
may still argue for keeping only a few terms, giving for instance a vev 
of order $\Mpl$. For field values one or more orders of magnitude bigger
there is no theoretical argument for such an assumption.
Equally there is none against it (except in the case of
the moduli fields of superstring theory).

Returning to the small-field case, not
all terms need be significant. In particular the 
$\phi^4$ term can be absent, or at least suppressed by factors like 
$(100\GeV/\Mpl)^2$ which make it completely negligible. If that happens,
$\phi$ is said to correspond to a flat direction in field space
(see for example Ref.~\cite{thermal} and references cited there).
As we shall note later a `flat' direction in this sense is a good 
candidate for ordinary hybrid inflation. (Note though
that this use of the word `flat' is not directly related to 
the `flatness conditions'
$\epsilon\ll1$ and $|\eta|\ll 1$ that are needed 
for slow-roll inflation {\em per se}.)

The above form of the expansion is appropriate if $\phi$ is the modulus
of a complex field charged under a $U(1)$ (or higher) symmetry,
the vevs of other fields charged under the symmetry being 
negligible. In general odd 
terms will appear, though the linear term
is still forbidden if the origin is the fixed point of some lower
symmetry, such as  a subgroup $Z_n$ of a $U(1)$.

In non-hybrid models, where only the inflaton field is relevant,
the above expansion is what we need, with $\phi$ 
the inflaton field. In hybrid inflation, where the dominant contribution 
to the potential comes from a different field $\psi$, we need a similar 
expansion in both fields simultaneously. 

If two or more fields are significant we have to worry about the fact
that in the context of supergravity one does not expect the fields to be 
canonically normalized. Rather, the kinetic term will be of the form
\be
{\cal L}_{\rm kin} = \frac12\sum_{a,b} h_{ab} \partial_\mu \phi_a
\partial^\mu \phi_b
\label{hmetric}
\ee
If the metric $h_{ab}$ has vanishing curvature then one can redefine the 
fields to get the canonical normalization $h_{ab}=\delta_{ab}$.
In supergravity the curvature does not vanish, but it will be negligible
in the small-field regime. For field values $\gsim\Mpl$ it will become
significant and in general must be taken into account. 
But even in a hybrid inflation model, the non-inflaton field is fixed 
(or nearly so) during inflation, and one can always
impose canonical normalization on a single-component inflaton field
even though that might not be the most natural choice in the context of 
supergravity.
Unless otherwise stated, the inflationary potential $V(\phi)$ is 
supposed to refer to a canonically normalized inflaton field
in what follows. 

In the above discussion we have in mind the case of Einstein gravity,
corresponding to ${\cal L}=-\frac12\Mpl^2 R + \tilde{\cal L}$. The 
second term is the Lagrangian defining the particle theory, which we 
have in mind is some gauge theory formulated in the usual context of 
essentially flat spacetime.
This is a reasonable assumption because the energy scale when 
cosmological scales leave the horizon during inflation
is at least two orders of magnitude below
the Planck scale. Modifications of Einstein gravity are sometimes 
considered though, such as replacing $\Mpl$ by a field or adding
an $R^2$ term. In these cases, and many others, one can recover
Einstein gravity by redefining the fields and 
the spacetime metric, at the expense of making the non-gravitational 
part  of the Lagrangian more complicated. In this way one typically arrives
at a model of inflation with large fields and a non-polynomial 
potential, such as the one in Eq.~(\ref{another}). Models of this kind are 
in practice constructed without reference to supergravity,
and ignore our lack of understanding of the large-field behaviour
of the potential just as much as large-field models invoking say a power-law
potential.

\subsection{The difficulty of keeping $\eta$ small in supergravity}

In small-field models of inflation, where the theory is under control,
it is reasonable to work in the context of 
supergravity. This is a relatively recent activity because although 
small-field models were the first to be proposed \cite{new,singlet} 
they were soon abandoned in favour of models with fields first of order
the Planck scale \cite{primordial} and then much bigger \cite{CHAOTIC}.
Activity began again after hybrid inflation was proposed
\cite{l90,LIN2SC}, with the realization
\cite{CLLSW} that the model is again of the small-field type.
In Ref.~\cite{CLLSW} supersymmetric implementations of 
hybrid inflation were considered, in the context of both global 
supersymmetry and of supergravity.

According to supergravity, the potential can be written as 
a `$D$ term' plus an `$F$ term', and it is usually supposed that
the $D$ term vanishes during inflation. 
All models proposed so far are of this kind, except for
hybrid inflation models with the $D$ term dominating
\cite{ewansgrav,bindvali,halyo}; such models are 
quite special and are expected to give a distinctive
$n=0.96$ to $0.98$ as briefly mentioned in
Section V.

For models where the $D$ term vanishes,
the flatness parameter $\eta=\Mpl^2 V''/V$ generically receives
various contributions of order $\pm 1$. This 
crucial point was first emphasized in Ref.~\cite{CLLSW}, though it
is essentially 
a special case of the more general result,
noted much earlier
\cite{dinefisch,coughlan}, that 
there are contributions of order $\pm H^2$ to the mass-squared of every
scalar field. Indeed, in a small-field
model the troublesome contributions to $\eta$ may
be regarded as contributions
to the coefficient $m^2$ in the expansion (\ref{power}) of the inflaton
potential. 

Since slow-roll inflation requires 
$|\eta|\ll 1$, either there is an accidental
cancellation, or the form of the $F$ term is non-generic.
An accident is reasonable only if $\eta$ is not too small, 
corresponding to $n=1+2\eta$ not too close to 1.

Pending a full discussion in Section VII, the $\eta$ problem 
will be mentioned as appropriate when discussing the various models.

\subsection{Before cosmological inflation}

The only 
era of inflation 
that is directly relevant for observation is the one beginning when 
the presently observable universe leaves the horizon.
This era of `cosmological inflation' 
will undoubtedly be preceded by
more inflation, but all memory of earlier inflation is lost apart from
the starting values of the fields at the beginning of cosmological 
inflation. Nevertheless, one ought to try to understand the earlier era
if only to check that the assumed starting values are not ridiculous.

A complete history of the universe will 
presumably start when  the energy density is at the Planck scale.
(Recall that $V^{1/4}$ is at least two orders of magnitude 
lower during cosmological  inflation.) The
generally accepted hypothesis is that the scalar fields
at that epoch take on chaotically varying values as one moves around
the universe, inflation occurring where they have suitable values
\cite{CHAOTIC,abook}.
It is indeed  desirable to  start descending from the Planck scale
  with an era of inflation
for at least two reasons. One is
to avoid the universe either collapsing or becoming practically empty,
and the other is to have an event horizon so that the
homogeneous patch within which we are supposed to live is
 not eaten up by its inhomogeneous
surroundings. However, there is no reason to suppose
that this initial era of inflation is of the slow-roll variety.
The motivation for slow-roll comes from the observed 
fact that $\delta_H$ is almost scale-independent, which applies only 
during the relatively brief era when cosmological scales are 
leaving the 
horizon. In the context of supergravity,
where achieving slow-roll inflation requires rather delicate 
conditions, it might be quite attractive to suppose that
non-slow-roll inflation takes the universe down from the Planck scale with 
slow-roll setting in only much later. A well known potential 
that can give non-slow-roll inflation is
$V\propto \exp(\sqrt{2/p}\phi/\Mpl)$, which gives
 $a\propto t^p$ and corresponds to non-slow-roll inflation in
the regime where $p$ is bigger than 1 but not much bigger.

If slow-roll is established well before cosmological scales leave the 
horizon it is possible to have an era of `eternal inflation'
during which
the motion of the inflaton field is dominated by the
quantum fluctuation.{For eternal inflation taking place at 
large field values see
Ref.~\cite{eternal}. The corresponding phenomenon for inflation near a 
maximum was noted earlier 
by a number of authors.}
The condition for this to occur is
that the predicted spectrum $\delta_H$ be 
formally bigger than 1 \cite{stochastic}.

Coming to the beginning of cosmological inflation, the 
possibilities are actually so varied 
that one can contemplate a wide range of initial field values.
Going back in time, one might find a smooth
inflationary trajectory going all the way back to an era when
$V$ is at the Planck scale 
(or at any rate much bigger than its value during cosmological 
inflation). In that case
the inflaton field will probably be decreasing during inflation.
Another natural possibility is for the inflaton to 
find itself near a maximum of the potential 
before cosmological inflation 
starts.
Then there may be eternal inflation followed by 
slow-roll 
inflation.
If the maximum is a fixed point of the symmetries it is quite natural for 
the field to have been driven there by its interaction with other 
fields. Otherwise it
could arrive there by accident, though this 
is perhaps only reasonable if
the distance from the maximum to the minimum
is $\gsim \Mpl$
(see for instance Ref.~\cite{natural3} for an example).
In this latter case, the fact that eternal inflation occurs near the 
maximum may help to enhance the probability of inflation starting 
there.\footnote
{The idea that 
inflation could start at a maximum of the potential through 
non-thermal effects was first promoted in Ref.~\cite{nontherm}.
If the maximum is a fixed point the field might alternatively
be placed there through thermal corrections to the potential
\cite{new}, but
this `new inflation'
hypothesis is difficult to realize and has fallen out of favour now that 
alternatives have been recognized.}

\section{Non-hybrid models}

Non-hybrid models of inflation proposed so far 
fall into two broad classes.
Either the inflaton field is of order $10\Mpl$ when cosmological scales 
leave the horizon, moving towards the origin
under a power-law potential $V\propto \phi^p$, or it is $\lsim \Mpl$
and moving away from the origin under a potential
$V\simeq V_0(1-\mu\phi^p)$. The first case is often called 
`chaotic' inflation because it provides a way of 
descending from the Planck scale with chaotic initial field conditions.
The second case is often called `new' inflation because that was the 
name given to the first example of it. 

\subsection{Power-law and exponential potentials}

The simplest potential giving inflation is 
\be
V= \frac12 m^2\phi^2
\ee
and obvious generalizations are $V=\frac14\lambda\phi^4$,
and $V=\lambda M_{\rm Pl}^{4-p}\phi^p$ with $p/2\geq 3$.
Potentials of this form were proposed as the simplest realizations
of chaotic initial conditions at the Planck scale
\cite{CHAOTIC}.

Inflation ends 
at $\phi_{\rm end}\simeq p\Mpl$, after which $\phi$
starts to oscillate about its vev $\phi=0$.
When cosmological scales leave the horizon
$\phi= \sqrt{2Np}\Mpl$. Since the inflaton field is
of order 1 to $10\Mpl$, there is no 
particle physics motivation for a power-law potential
though equally
there is no argument against it.

The model gives  $n-1=-(2+p)/(2N)$, and gravitational waves 
are big enough to be 
eventually observable with $r=2.5p/N=5(1-n)-2.5/N$. 
The COBE normalization 
corresponds to $m=1.8\times 10^{13}\GeV$ for the quadratic case.
For $p=4,6,8$ it gives respectively
$\lambda=2\times 10^{-14}$, $\lambda=8\times 10^{-17}$,
$\lambda=6\times 10^{-20}$ and so on.
The energy scale is $V^{1/4}\sim 10^{16}\GeV$ (corresponding to
$\epsilon$ not very small)
but as we are not dealing with a broken symmetry there is no
question of relating this to a GUT scale.

The same prediction is obtained for a more complicated potential,
provided that it is proportional to $\phi^p$ during cosmological
inflation,
and in particular $\phi$ could have a nonzero vev $\ll\Mpl$
\cite{shafi1,kls}. The more general case, where more than one power of
$\phi$ is important even while cosmological scales are leaving the 
horizon, has also been considered \cite{hodges}. 
It can give rise to 
a variety of predictions \cite{blumhodges}
for the scale-dependence of $\delta_H$
but of course requires a delicate balance of coefficients. 

The limit of a high power is  an exponential potential,
of the form $V=\exp(\sqrt{2/q}\phi)$. This gives
 $\epsilon=\eta/2=1/q$
which lead to
$n-1=-2/q$ and $r=10/q$.
This is the case of 
`extended inflation',
where the basic interaction involves non-Einstein gravity but the
exponential potential occurs after transforming to Einstein gravity
\cite{extended}.
However, simple versions of this proposal are ruled out by observation,
because  the end of inflation corresponds to a first order phase
transition, and in order for the bubbles not to spoil the cmb
isotropy one requires $n\lsim 0.75$ in contradiction with
observation \cite{LL1,green}.

\subsection{The inverted quadratic potential}

Another simple potential leading to inflation is
\cite{nontherm,bingall,natural3,paul,natural,ovrut,natural2,%
natural4,kinney,kumekawa,izawa}
\be
V=V_0-\frac12m^2\phi^2 +\cdots
\label{natural}
\ee
We shall call this the `inverted' quadratic potential, to distinguish it 
from the same potential with the plus sign which comes from the simplest 
version of hybrid inflation.
The dots indicate the effect of higher powers, that are supposed 
to come in after cosmological scales leave the horizon. As long as they 
are negligible, $V_0$ has to dominate for inflation to 
occur. 

This potential gives $1-n=2\Mpl^2m^2/V_0$. If $m$ and $V_0$ are regarded 
as free parameters, the region of parameter space permitting slow-roll
inflation corresponds to $1-n\ll 1$. Thus $n$ is indistinguishable from
1 except on the edge of parameter space. However, in the context of 
supergravity there are two reasons why the edge might be regarded as 
favoured. One is the fact that 
$\eta$ generically receives contributions of order
1. Since slow-roll inflation 
requires $|\eta|\ll 1$, either $\eta$ is somewhat reduced from its 
natural value by accident, or it is suppressed because the theory has a 
non-generic form. In the first case one does not expect $\eta$ to be 
tiny, and in the present context this means that $1-n=2\eta$ will not be 
tiny either.

The other reason for expecting $n$ to be significantly below 1, which is 
specific to this potential, has to do with the magnitude of the vev
$\phi_m$. If the inverted quadratic form
for the potential holds until $V_0$ ceases to dominate, 
one expects
\be
\phi_m\sim \frac{V_0^{1/2}}{m}
=\left(\frac{2}{1-n} \right)^{1/2} \Mpl 
\ee
(This is also an estimate of $\phi_{\rm end}$ in that case.)
To understand the potential within the  context of particle
theory the vev should not be more than a few times
$\Mpl$, which requires $n$ to be well below 1. 

We see that to have $n$ indistinguishable from 1, the
potential should steepen drastically after cosmological scales leave the 
horizon. Even given that condition, one also requires that the smallness
of $\eta$ is not an accident in the context of supergravity.
So far no model has been proposed that satisfies both of these 
conditions, though a steepening of the potential with an accidental 
suppression of $\eta$ has recently been considered \cite{izawa}.

In the next section we shall also consider the inverted quadratic
potential in the context of hybrid inflation, where some other field is 
responsible for $V_0$. In that case inflation ends at
some $\phi_{\rm end}\ll \Mpl$, and one can be dealing with a version of 
supergravity that keeps $\eta$ tiny in which case
$n$ will be indistinguishable from 1.

When cosmological scales leave the horizon
$\phi\sim \phi_{\rm end} e^{-x}$ where $x\equiv N(1-n)/2$.
The COBE normalization gives
\be
\frac{V_0^{1/2}}{\Mpl^2}= 5.3\times 10^{-4} 
N^{-1} x e^{-x}\frac{\phi_{\rm end}}{\Mpl}
\ee
For $\phi_{\rm end}\sim \Mpl$ this gives $V_0^{1/4}> 5\times 10^{14}\GeV$
(using $n>0.7$). But in a hybrid inflation model $\phi_{\rm end}$
and $V_0^{1/4}$ could be much lower.

Staying with non-hybrid models, there have been several
proposals for realizing the inverted quadratic potential
without steepening. Since there is no steepening they 
will all give $n$ significantly below 1, and are not expected to
occupy the regime $0.9\lsim n<1$ populated by models to be discussed
later. Let us list them briefly.

\subsubsection{Modular inflation}

If $\phi$ is the real or imaginary part of 
one of the superstring moduli, its potential is generally expected
\cite{bingall,paul} to be of the form
$V=\Lambda^4 f(\phi/\Mpl)$, with $f(x) $ and its derivatives roughly 
of order 1 
in the regime $|x|\lsim 1$. 
In that region one expects the flatness parameters
$\eta\equiv\Mpl^2V''/V $
and $\epsilon \equiv \Mpl^2(V'/V)^2/2$ to be both roughly of order 1, 
and one might hope to find within it a region where they both happen to
be  significantly
below 1 so that slow-roll inflation can occur.
In the regime $|x|\gg 1$ one expects
either a potential that rises too steeply to support inflation,
or one that becomes periodic.

It is usually supposed that inflation takes place near a maximum.
(As the maximum is not usually supposed to be a fixed point
of a symmetry there is no reason why the interaction of the modulus 
should drive it there, but the existence of `eternal'
inflation may still be held to favour it as a starting point.)
One then expects
the potential to be of the inverted quadratic form, with
no significant steepening after cosmological scales leave the 
horizon.
So far, investigations using specific models \cite{bingall,natural2,macorra}
have actually concluded that viable inflation does not occur.\footnote
{Ref.~\cite{bento} claims to have been successful, but 
an analytic calculation of that model reported in Ref.~\cite{ournew} 
finds that it is not viable.}

Many complex fields other than the modulus are expected
in the context of supersymmetry, and one could take 
$\phi$ to be the real or imaginary part of one of them.
Although the potential for such fields is not understood for 
$\phi\gg\Mpl$ it is not unreasonable to suppose that there is a 
vev at $\phi\sim \Mpl$, and if so one can again expect an inverted 
quadratic form for the potential.

\subsubsection{`Natural' inflation}

The prediction for $n$ depends only on the form of the potential while
cosmological scales are leaving the horizon. A potential
reducing to the inverted quadratic one in this regime is
\be
V= V_0 \cos^2(\phi/M)
\label{gbpot}
\ee
This potential typically arises if $\phi$ is the pseudo-Goldstone boson
of
a $U(1)$ symmetry broken by non-perturbative effects.
Inflation using it 
was first discussed
in Ref.~\cite{bingall}, and then in Ref.~\cite{natural} where it was called
`natural' inflation, and has subsequently been considered by several 
authors \cite{natural3,ovrut,natural2,natural4,glw,ewanprep}.

Suggestions as to the identity of $\phi$ have been that
it is the imaginary part of a modulus
such as the dilaton \cite{bingall,natural2}, 
the angular part
of an ordinary complex field 
\cite{natural,natural2,natural3,natural4,glw,ewanprep}
or something else \cite{ovrut}. For any of them to be reasonable
$M$ should be not too far above the Planck scale, corresponding
to $\eta$ not too small and $n$ significantly below 1.

Even when $\eta$ is not tiny, one still feels more comfortable with a 
definite mechanism for keeping it well below 1 in the context of 
supergravity, and the $U(1)$ symmetry is such a mechanism
provided that it is broken only by the superpotential
(Section VII).

\subsubsection{A loop-corrected potential}

It was pointed out in Ref.~\cite{ewanloop} that a
different mechanism for keeping $\eta$ small 
might be provided by a loop correction of the 
form $A\phi^2\ln(\phi/B)$ \cite{ewanloop}.
Since the loop correction has to be large one needs to allow a renormalization 
group modification of it. The most reasonable model to 
emerge has an inverted quadratic potential without steepening,
leading again to $n$ well below 1.

\subsection{Cubic and higher potentials}

If the quadratic term is heavily suppressed or absent, one will have
\be
V\simeq V_0 ( 1-\mu \phi^p +\cdots)
\label{higher}
\ee
with $p\geq 3$. For this potential one expects that the integral
(\ref{nint}) for $N$ is dominated by the limit
$\phi$ leading to \cite{kinney} 
\be
\phi^{p-2}=[p(p-2)\mu N \Mpl^2]^{-1}
\label{phiend}
\ee
and
\be
n\simeq1-2\left(\frac{p-1}{p-2}\right)\frac1 N
\ee

It is easy to see that the integral is typically dominated by the $\phi$ limit,
if higher terms in the potential (\ref{higher}) become significant only when
$V_0$ ceases to dominate at 
$\phi^p\sim \mu^{-1}$. Then, in 
the regime where $V_0$ dominates,
$\eta=[(p(p-1)\Mpl^2/\phi^2]\mu\phi^p$, and if this expression becomes of order
1 in that regime inflation presumably ends soon after. 
Otherwise inflation ends when $V_0$ ceases to 
dominate. At the end of inflation one therefore has 
$\Mpl^2\mu\phi_{\rm end}^{p-2}\sim 1$ if $\phi_{\rm end}\ll 
\Mpl$, otherwise one has $\mu\phi_{\rm end}^p\sim 1$.
(We are supposing for simplicity that $p$ is not enormous, and dropping 
it in these rough estimates.)
The integral
(\ref{nint}) is dominated by the limit
$\phi$ provided that 
\be 
N\Mpl^2\mu\phi_{\rm end}^{p-2}\gg 1
\label{criterion1}
\ee
This is always
satisfied in the first case, and is satisfied in the second case
provided that $\phi_{\rm end}\ll \sqrt N \Mpl$ which we shall assume.
If higher order terms come in more quickly than we have supposed, or if
inflation ends through a hybrid inflation mechanism then
$\phi_{\rm end}$ will be smaller than these estimates, and 
one will have to see whether the criterion (\ref{criterion1})
is satisfied.

Continuing with the assumption that it is satisfied, 
the COBE normalization is \cite{kinney} 
\be
5.3\times 10^{-4}= (p\mu\Mpl^p)^{\frac1{p-2}} [N(p-2)]^{\frac{p-1}{p-2}}
V_0^{\frac12} \Mpl^{-2}
\ee
In particular, for $p=4$ the dimensionless coupling is
$\lambda\equiv 4V_0\mu=2.8\times 10
^{-13}(50/N)^3$. At least in the context of non-hybrid models 
such a tiny value may be problematical, as 
was recognized in the first models of inflation
\cite{new,singlet} using the essentially equivalent non-supersymmetric 
Coleman-Weinberg loop correction $A\phi^4(\phi/B)$
(see however Ref.~\cite{langbein} for a dissenting view about this
latter case).

As we noted earlier, a strong suppression of the quadratic term 
may well occur in a supergravity theory.
An explicit example has been given recently
\cite{grahamnew}. In this proposal, one finds a potential
\be
V=V_0(1+\beta\phi^2\psi - \gamma\phi^3 +\cdots)
\ee
where $\psi$ is another field. Then, with $\beta$ and $\gamma$ of order 1 in 
Planck units, and initial values 
$\psi\sim \Mpl$ and $\phi\simeq 0$ one can check that
the quadratic term is driven to a negligible value
before cosmological inflation 
begins.

This proposal gives $\phi$ a vev of order $\Mpl$.
Some particle-physics motivation for 
small-field models with $p\geq 3$ is given in
 Refs.~\cite{natural4,kinney}, though not in the context of
supergravity.

One could contemplate models in which
more than one power of $\phi$ is significant while 
cosmological scales leave the horizon, but this requires a delicate 
balance of coefficients. Models of this kind were also discussed a long 
time ago \cite{primordial,olive}, again with a vev of order $\Mpl$,
but their motivation was in the context of setting the initial value of 
$\phi$ through thermal equilibrium and has disappeared with the 
realization that this `new inflation' mechanism is not needed.
They could give a range of predictions for $n$.

\subsection{Another potential}
                    
Potentials have been proposed
that are of the form
\be
V\simeq V_0(1- e^{-q\phi/\Mpl})
\label{another}
\ee
with  $q$ of order 1.
This form is supposed to 
apply in the regime where $V_0$ dominates, which is $\phi\gsim \Mpl$.
Inflation ends at $\phi_{\rm end}\sim \Mpl$, and when 
cosmological scales leave the horizon one has
\bea
\phi&=& \frac1q \ln(q^2 N) \Mpl\\
n-1&=&-2\eta= -2/N 
\label{r2pot}
\eea
Gravitational waves are negligible.

The most reasonable-looking derivation of a potential of this form
in the context of particle theory
\cite{ewansgrav} starts with a highly non-minimal kinetic term.
In that case $q$ can have different values such as
                    $1$ or $\sqrt2$.
Another derivation \cite{r2,burt} modifies Einstein gravity
by adding a large $R^2$ term to the usual $R$ term, but with a 
huge
coefficient,
and a third \cite{vplanck} uses a variable Planck mass. In both cases,
after transforming back to Einstein gravity one obtains the above form
with $q=\sqrt{2/3}$.

This potential is mimicked by $V=V_0(1-\mu \phi^{-p})$ with
$p\to\infty$ (Table 1).

\section{Hybrid inflation models}

In hybrid inflation models \cite{l90,LIN2SC}, the 
slowly rolling inflaton field $\phi$ is 
not the one responsible for most of the energy density.\footnote
{We take this to be the definition of the term `hybrid', which is 
coming to be the standard usage though somewhat narrower than 
in Refs.~\cite{LIN2SC,LIN2SC2} where it was first introduced.}
That role is 
played by another field $\psi$, which is held in place
by its interaction with $\phi$ until the latter reaches a critical
value $\phi_c$. When that happens $\psi$ is destabilized
and inflation ends. As was pointed out in Ref.~\cite{CLLSW},
$\phi_c$ is typically small in Planck units which means that hybrid 
inflation models are typically small-field models.

The end of inflation can correspond to either a first \cite{l90}
or a second order \cite{LIN2SC} 
phase transition, but the second-order case (corresponding to the absence of a 
potential barrier) is simpler and is the only one that will be 
considered here.
It has been considered by many authors
\cite{LL2,LIN2SC2,CLLSW,ewansgrav,lisa,glw,ewanloop,ournew,%
mutated,bindvali,halyo,ewanprep,%
silvia,wang,dave,qaisar,dvaliloop,lazpan,qaisarlatest}.
(The first-order case amounts to an
Einstein gravity version of extended inflation \cite{extended,green}, 
and has been considered in Refs.~\cite{AdamsFreese,CLLSW}.
For a non-Einstein version of second-order hybrid inflation see
Ref.~\cite{bertolami}.)

\subsection{Ordinary hybrid inflation}

The potential for the original model of (second-order) hybrid inflation 
is \cite{LIN2SC}
\bea
V&=&\frac14\lambda(M^2-\psi^2)^2 +\frac12m^2\phi^2
+\frac12\lambda'\psi^2\phi^2 \\
&=&V_0-\frac12 m_\psi^2 \psi^2 + \frac14\lambda\psi^4 
 + \frac12 m^2 \phi^2 +\frac12\lambda'\psi^2\phi^2 
\label{fullpot}
\eea
The field $\psi$ is fixed at the origin if $\phi>\phi_c$, where
\be
\phi_c^2=m_\psi^2/\lambda'=\lambda M^2/\lambda'
\ee
In this regime slow-roll inflation can take place, with
the quadratic potential 
\be
V=V_0+ \frac12 m^2\phi^2
\label{vord}
\ee

In the context of particle theory one expects the 
couplings $\lambda$ and $\lambda'$ to be $\lesssim 1$.
In order to have inflation at small $\phi$
the first term must dominate after cosmological 
scales leave the horizon, because at the other extreme one arrives at
$V\propto \phi^p$ which requires $\phi\sim 10$ to $20\Mpl$.
To fully justify the low-order polynomial form of the potential
one would also like to have $M\ll\Mpl$, which also
turns out to be a requirement 
for inflation to end promptly at $\phi_c$ \cite{CLLSW}.
It is reasonable to also consider $M\sim \Mpl$, which has been done
in Refs.~\cite{lisa,glw}.

The same inflationary potential is obtained with other forms for the 
last term of the potential (\ref{fullpot}), an example being 
\cite{lisa}
$\lambda'\Mpl^{-2} \psi^2\phi^4$, leading to 
\be
\phi_c^2=\Mpl m_\psi/\lambda'=\Mpl M \lambda^{1/2}/\lambda'
\label{phic2}
\ee
Other prescriptions for 
$\phi_c$ are provided by mutated hybrid inflation as described
later.

When cosmological scales leave the horizon
\be
\frac{\phi}{\phi_c}=e^{\frac{n-1}{2}N}
\label{phih}
\ee
At least  with the above two prescriptions 
for $\phi_c$ 
this implies that we are dealing with a model 
in which the inflaton field is small \cite{CLLSW}.

The quadratic inflationary potential gives \cite{LL2}
$n=1+2\Mpl^2 m^2/V_0$. From
the flatness condition $\eta\ll 1$, the 
allowed region of parameter space corresponds to $n-1\ll 1$, so 
one expects $n$ to be indistinguishable from 1 unless there is some
reason for the parameters to be near the edge of the allowed region.
Two reasons might be cited for wanting to be near the edge, but neither
is very compelling. One would be the fact that in 
 $\eta\equiv \Mpl^2 V''/V$ is of order 1 in
a {\em generic} supergravity theory.
If the smallness of $\eta$ is due to an accident it should not be too small.
However, in contrast with the inverted quadratic potential discussed 
earlier, we are now dealing with a model in which the inflaton field is 
small. As discussed in Section VII this makes it more attractive
to suppose that the smallness of $\eta$ is ensured from the start,
in which case there is no reason why it should not be tiny.
A more subtle consideration 
is the proposal of Ref.~\cite{lisa} that
$m\sim m_\psi \sim 100\GeV$ and $M\sim \Mpl$ are favoured values, which
taken literally 
does indeed give $\eta\sim 1$.
However the COBE normalization, to be discussed 
in a moment, would then require a precise 
choice of $\phi_c$, and in particular with either of the
prescriptions one would need a precise value of $\lambda'$ which is 
{\em not} of order 1. Thus, while the suggested orders of magnitude
 may be reasonable as rough estimates,
it cannot be said that the extreme edge of 
the allowed region is really favoured, and indeed of the six examples 
displayed in the Figures of Ref.~\cite{lisa} all except one
have $n$ indistinguishable from 1.\footnote
{Since the actual models used in Ref.~\cite{lisa}
are formulated in the context of global supersymmetry the smallness of 
$\eta$ should presumably be viewed as an accident for them.
But one can imagine that a different 
implementation of the proposal might keep $\eta$ small automatically
and we have proceeded with the discussion on this assumption.
If the smallness of $\eta$ is really accidental, very small values 
are disfavoured and one should reject the other five examples
in favour of the last one. 
This point seems to have been overlooked in Ref.~\cite{lisa}.}

The conclusion is that in the ordinary hybrid inflation model
one expects $n$ to be 
indistinguishable from 1, though a value significantly 
above 1 is not out of the question.

The COBE normalization is
\be
5.3\times 10^{-4}=\Mpl^{-3}\frac{V_0^{3/2}}{m^2\phi}
\label{cobenormhyb}
\ee
where $\phi$ is given by Eq.~(\ref{phih}).
With $\phi_c$ given
by either of the above prescriptions this imposes \cite{CLLSW}
a limit
$n\lesssim 1.3$ (assuming that 
$V_0$ dominates the potential and that $M\lsim \Mpl$).

What about higher powers of $\phi$ in the inflationary potential?
They can be ignored if their contribution is 
insignificant when $\phi$ has the value given by (\ref{cobenormhyb}).
For a quartic term $\frac14\lambda\phi^4$ 
this requires $\lambda\ll 10^{-7}(1-n)^3$ which seems very small.
However, the quartic term can be 
eliminated altogether, 
by identifying $\phi$ with one of the `flat' directions of 
particle theory.

If a higher power $\phi^p$ does dominate
we have
\be
n=1 + 2\left(\frac{p-1}{p-2}\right)\frac1 N
\ee
For $p=4$ the COBE normalization requires
\cite{dave}
a very small dimensionless coupling
$\lambda\equiv4\mu V_0\lsim 10^{-12}$. This case is therefore less 
attractive than the $p=2$ case. With higher powers,
inflation can
be interrupted and the requirement that the interruption is
over before cosmological scales leave the horizon
places an additional restriction
on the parameter space \cite{dave}.
The case where quadratic, cubic and quartic terms may all be important 
has also been discussed \cite{wang}, though the possibility of inflation 
being interrupted was ignored. As with the example discussed at the end 
of the next subsection (and the large-field model of 
Ref.~\cite{hodges}) there are enough parameters to allow a variety of
possibilities for the scale dependence of $\delta_H(k)$, though a 
delicate balance between the parameters is required to achieve this.
A linear term
$p=1$ is forbidden if the origin is a fixed point of appropriate 
symmetries, and even if a linear term is present the case that it 
dominates is somewhat artificial (see the example at the end of 
the next subsection). For the record, it gives a 
spectral index indistinguishable from 1 in a 
small-field model, because $V''=0$. 

To summarize, ordinary hybrid inflation leading to a a quadratic inflationary
potential $V(\phi)$ is a very attractive model. The full potential
$V(\phi,\psi)$ need contain only quadratic and
quartic terms, and need have no small couplings after the quartic term
in $\phi$ has been eliminated by identifying $\phi$ with a `flat' direction.
Of course, whether the model can be successfully embedded in a more 
complete theory is a bigger question, which is not the subject of 
the present work, and is still a long way from being answered.

\subsection{Inverted hybrid inflation}

Instead of $\phi$ rolling towards the origin it might roll away from it.
The simplest way of achieving this `inverted' hybrid inflation 
\cite{ournew} is to have\footnote
{As was pointed out a long time ago \cite{weinberg}, a potential of this kind
with the fields in thermal equilibrium 
leads to high temperature symmetry restoration, the inverse of the usual 
case. The same thing is happening in our non-equilibrium situation.}
\be
V = V_0 - \frac{1}{2} m_\phi^2 \phi^2 + \frac{1}{2} m_\psi^2 \psi^2
- \frac{1}{2} \lambda \phi^2 \psi^2 + \frac14\lambda_\phi\phi^4
+\frac14\lambda_\psi\psi^4
\label{first}
\ee
There is supposed to be a minimum at nonzero $\phi$ and $\psi$,
at which $V$ vanishes, but at fixed $\phi$ there is a minimum
at $\psi=0$ provided that
\be
\phi < \phi_{\rm c} = \frac{m_\psi}{\sqrt{\lambda}}
\ee
In this regime one can have inflation with the inverted quadratic potential
\be
V=V_0-\frac12m^2\phi^2
\ee
This is the same potential that we discussed already, with
the hybrid inflation mechanism now ending inflation while the field is still 
small, instead of the steepening of the potential that has to be 
postulated in a non-hybrid model. The hybrid mechanism looks more 
natural, because it involves only 
quadratic and quartic terms, with no need for the dimensionless 
couplings of quartic terms to be small.

One 
could replace the quadratic term $-\frac12m^2\phi^2$ by 
a higher order term, but the hybrid inflation mechanism would not then
offer any simplification compared with the corresponding 
single-field models.
If two or more terms are comparable while cosmological scales leave 
the horizon, there are more complicated 
possibilities including a significant 
variation of the spectral index. 
For example, one could have
\be
V=V_0-\frac12m^2\phi^2+\frac14\lambda\phi^4
\ee
This has a minimum at $\phi_m=m/\sqrt\lambda$ and we suppose that 
$V_0$ still dominates there. Then, if
$\phi_c\sim \phi_m$ the spectral index could flip between the values
$n=1\pm2\Mpl^2m^2/V_0$ on cosmological scales.
The number of $e$-folds taken to flip is $\Delta N\sim
\Mpl^{-2}(V/V')\phi_m \sim (\Delta n)^{-1}$ where
$\Delta n$ is the magnitude of $1-n$ before and after the
flip. For the flip to occur with $\Delta N<10$ (the range of 
cosmological scales) requires $\Delta n\gtrsim 0.1$.

For $\Delta n\lsim .1$, 
the whole of cosmological inflation could 
take place with $\phi$ midway between the maximum and minimum. In 
particular it could be near the point of inflexion so that the potential 
is practically linear. By choosing the same point as the origin we 
recover the case $p=1$ discussed earlier. We see that at least in this 
particular realization that case is either unnatural because the
cosmological epoch has to occupy a special place on the trajectory,
or is trivial because the entire trajectory (not just its linear part)
gives $n$ indistinguishable from 1. 

\subsection{Mutated hybrid inflation}

In both ordinary and inverted hybrid inflation, the other field $\psi$ is 
precisely fixed during inflation. If it varies,
an effective potential $V(\phi)$ can be generated even if the
original potential contains no piece that depends only on $\phi$.
This mechanism was first proposed in Ref.~\cite{mutated}, where it was called 
mutated hybrid inflation. The potential considered was
\be
V = V_0 - A\psi + B\psi^2 \phi^2 + C\psi^2
\ee
The last term serves only to give $V$ a minimum at which it vanishes, 
and is negligible during inflation. All of the other terms are significant,
with $V_0$ dominating.
For suitable choices of the parameters inflation takes place
with $\psi$ held at the instantaneous minimum, leading to a potential
\be
V= V_0 (1-\mu \phi^{-2})
\ee
Shortly afterwards the mechanism was rediscovered \cite{lazpan} and called
`smooth' hybrid 
inflation emphasizing that any topological defects associated with 
$\psi$ will never be produced
(in contrast with the case of ordinary and inverted hybrid inflation).
The potential considered there was $V=V_0-A\psi^4+B\psi^6\phi^2
+C\psi^8$, where again the last term is negligible during inflation
and $V_0$ dominates the remainder. It leads to 
$V=V_0(1-\mu \phi^{-4})$.

Retaining the original name, the most general mutated hybrid inflation 
model with only two significant terms is \cite{ournew}
\be
V = V_0 - \frac{\sigma}{p} \Mpl^{4-p}\psi^p + \frac{\lambda}{q} 
\Mpl^{4-q-r}\psi^q \phi^r
+\dots
\ee
In a suitable regime of parameter space,
$\psi$ adjusts itself to minimize $V$ at fixed
$\phi$, and
$\psi\ll\phi$ so that the slight curvature of the 
inflaton trajectory does not affect the field dynamics. Then,
provided that $V_0$ dominates the energy density, the effective potential 
during inflation is
\be
V=V_0(1-\mu \phi^{-\alpha})
\label{vmut}
\ee
where
\bea
\mu&=&\Mpl^{4+\alpha} \left(\frac{q-p}{pq}\right)
\frac{\sigma^{\frac q{q-p}}\lambda^{-\frac{p}{q-p}}}
{V_0} >0\\
\alpha&=&\frac{pr}{q-p}
\eea
For $q>p$, the exponent $\alpha$ is positive as in the examples already 
mentioned, but for $p>q$ it is negative with $\alpha<-1$.
In both cases it can be non-integral, though integer values are the most 
common for low choices of the integers $p$ and $q$.

The situation in 
the regime $-2\lsim\alpha<-1$ is similar to the one that we 
discussed already for the case $\alpha=-2$;
the prediction for $n$ covers a continuous range below 1 because it
depends on the parameters,
but to have a small-field model the potential has to be steepened
after cosmological scales leave the horizon. An example of such 
steepening is provided in the next subsection.

For choices of $\alpha$ outside the range $-2\lsim\alpha<0$,
the integral (\ref{nint}) is dominated by the limit $\phi$
provided that we are dealing with a small-field model, corresponding to
\be
\phi_{\rm end}\simeq \mu^{1/\alpha}\ll \Mpl
\label{phiend2}
\ee
Then one has \cite{ournew}
\be
n = 1 - 2 \left(\frac{\alpha+1}{\alpha+2} 
\right)\frac{1}{N}
\ee
This prediction is listed in the Table for some integer values of 
$\alpha$, along 
with the limiting cases $\alpha=\pm\infty$ and $\alpha=0$.

Of the various possibilities regarding $\alpha$, some 
are preferred over others in the context of supersymmetry.
One would prefer \cite{ournew} $q$ and $r$ to be even if
$\alpha>0$ (corresponding to $q>p$)
and $p$ to be even if $\alpha<0$. Applying this criterion with $p=1
$ or $2$ and $q$ and $r$ as low as possible leads \cite{ournew}
to the original mutated hybrid model, along with the cases
$\alpha=-2$ and $\alpha=-4$ that we discussed earlier
in the context of inverted hybrid and single-field models.
In the original model (at least)
the order of magnitude of the inflationary energy scale
$V_0^{1/4}$ can be understood if supersymmetry is broken
by gaugino condensation in a hidden sector.

A different example of a mutated hybrid inflation potential is given in
Ref.~\cite{glw}, where $\psi$ is a pseudo-Golstone boson
with the potential (\ref{gbpot}). Depending on the parameter values
it might reduce in practice to a potential of the form discussed above,
to one of the kind discussed in the next subsection or to something 
different.

\subsection{Mutated hybrid inflation with explicit $\phi$ dependence}

So far we have assumed that the original potential has no piece that
depends only on $\phi$. If there is such a piece it has to be added 
to the inflationary potential (\ref{vmut}). If it 
dominates while cosmological scales leave the horizon, 
the only effect that the $\psi$ variation  has 
on the inflationary prediction is to 
determine $\phi_c$ through Eq.~(\ref{phiend2}). 

\subsection{Hybrid inflation with a loop-corrected potential}

In the 
context of supersymmetry one expects the correction to be
of the form $A\phi^2\ln(\phi/B)$ or $A\ln(\phi/B)$,
depending on the mechanism of supersymmetry breaking during inflation.
We discussed an application of the first case earlier, and now consider 
the second. 

The prediction for $n$ with a loop correction of the form 
$A\ln(\phi/B)$ is the same 
as for a potential $V_0(1-\mu\phi^{-\alpha})$ with $\alpha\simeq 0$,
and as in the Table it is between .96 and .98.

It was first proposed \cite{qaisar,dvaliloop,qaisarlatest}
to give a slope to the flat classical potential coming out of a
globally supersymmetric model that had been written down 
in Ref.~\cite{CLLSW}. The superpotential is
\be
W=\sigma (\Phi_1\Phi_2 + \Lambda^2) \Phi_3
\label{wglobal}
\ee
where $\sigma$ is a dimensionless coupling, $\Lambda$ is a mass scale
and the
$\Phi_n$ are complex fields. This gives the classical potential
\be
V=\sum_n\left|\frac{\partial W}{\partial \Phi_n}\right|^2
\label{vglobal}
\ee
With $\Phi_1=\Phi_2=0$, $V$ 
has the constant value $\sigma^2\Lambda^4$, and the loop correction
gives it a small slope making $|\Phi_3|$ the inflaton of a hybrid 
inflation model. 

As had been pointed 
out in the earlier reference, one 
expects that in this model supergravity corrections will give
$\eta\equiv\Mpl^2|V''/V|\sim 1$, preventing inflation.\footnote
{Sufficient accidental suppression of $\eta$ is unlikely because the 
loop correction gives $\eta=.01$ to $.02$, and $W$ is not of a form that would 
guarantee the smallness of $\eta$ for a reasonable Kahler potential.}
However, it has 
recently been noted \cite{bindvali,halyo} 
that a loop correction of the same form is likely to 
be responsible for the slope of a hybrid inflation model dominated by 
the $D$ term, of the type that was proposed at tree level
in Ref.~\cite{ewansgrav}. As pointed out in that reference, such a 
$D$ term model 
has the nice feature that there is no problem
about keeping
$\eta$ small.

\section{The difficulty of inflation model-building in supergravity}

In this section the problem of keeping $\eta$ small in the context of 
supergravity is explained, and various proposed solutions to it are listed.
Most of them have been encountered already.

\subsection{The problem}

The
 kinetic terms of the complex
scalar fields
$\Phi_n$ are given in terms of the Kahler potential $K$
(a real function of $\Phi_n$ and its complex conjugate
$\bar\Phi_n$)
 by
\be
{\cal L}_{\rm kin} = (\partial_\mu\bar\Phi_n)
K_{\bar n m}
(\partial^\mu\Phi_m)
\ee
Here and in the following expressions, a subscript $n$
denotes the derivative with respect to the $\Phi_n$
($\bar n$ the derivative with respect to $\bar\Phi_n$)
and a summation over repeated indices is implied.
Following Ref.~\cite{ewansgrav}, 
focus on a given point on the inflationary trajectory,
and choose it as the origin $\Phi_n=0$. 
For the analysis leading to the usual slow-roll predictions to be valid
we need the relevant fields to be canonically normalized
in a small region around this point, which is ensured by choosing
the fields so that\footnote
{Strictly speaking this choice ensures
only that equations involving at most second derivatives of the fields
are valid. It ensures the validity of the prediction for
$\delta_H$ but not, strictly speaking, of the prediction for $n$.
The correction analogous to $\gamma$ in 
Eqs.~(\ref{deltan2}) and (\ref{running}) will
involve both $V'''$ and the higher order contributions
to $K_{\bar n m}$. It could be worked out in a particular model
from the general formalism of 
Ref.~\cite{nak}, specialized to a single field, and we are working on 
the assumption that it is negligible.}
\be
K_{\bar n m}=\delta_{nm}+O(\Phi_n^2)
\label{kinetic}
\ee
This is analogous to using a locally inertial frame in general
relativity.
The `curvature scale', beyond which higher order terms become
significant,
is expected to be of order the  Planck
scale $|\Phi_n|\sim \Mpl$. 

The potential $V$ consists
of a `$D$' term and an `$F$' term, and 
the problem arises when the $F$ term dominates, which
is usually taken to be the case. 
The $F$ term involves the superpotential $W$ (a  holomorphic function
of the complex fields $\Phi_n$) and  is
\be
V=e^{K/\Mpl^2}\tilde V
\label{vexp}
\ee
where
\be
\tilde V=(W_n+\Mpl^{-2}WK_n)K^{n \bar m} (\bar W_{\bar m} + 
\Mpl^{-2}\bar W K_{\bar m})
-3\Mpl^{-2}|W|^2
\ee
The matrix $K^{n \bar m}$ is the inverse of $K_{\bar n m}$.

This expression depends on $K$ and $W$ only through the combination
$G\equiv K+\ln|W|^2$, so it is invariant under the transformation
$K\to K-F-\bar F$, $W\to e^F W$ where $F$ is any holomorphic function. 
As a result one can choose 
$K$ and $W$ so that about the given point on the trajectory
\be
K=\sum_n|\Phi_n|^2+\cdots
\label{kahlerexp}
\ee
The inflaton field $\phi$ can be chosen to be $2^{-\frac12}$ times  the 
real part of one of the $\Phi_n$, and one then finds
 \be
\eta\equiv\Mpl^2V''/V=1+\Mpl^2\tilde V''/\tilde V
\ee
The first term, which alone would give $\eta=1$,
comes from the factor $e^{K/\Mpl^2}$. To
have $|\eta|\ll1$ it must be cancelled by the 
second term.
Examination of $\tilde V$ shows that there are
indeed contributions 
of order 1, which may be positive or negative,
but generically their total will not be $-1$ to high 
accuracy. They come from the quadratic term
in the expansion (\ref{kahlerexp}) of $K$, and \cite{ewanpers}
from the quartic $|\Phi|^4$ term through
$K^{\bar n m}$. Thus
the minimal supergravity 
approximation, of keeping only the quadratic term, 
cannot be used in this context.

\subsection{Solving the problem}

How severe the problem is depends on the magnitude of $\eta$.
If $\eta$ is 
not too small then its smallness could be due to accidental 
cancellations. Having $\eta$ not too small requires 
that $n-1=-6\epsilon+2\eta$ 
be not too small (unless $3\epsilon\simeq\eta$ which is not the case 
in models that have been proposed)
so the observational bound
$|n-1|<.3$ is already beginning to make an accident look unlikely.
An accidental cancellation is 
being assumed by proponents of 
modular inflation as discussed earlier.

A toy model for accidental cancellation would be the superpotential
$W=A(\Phi-\Phi_0)^2$. The inflaton is supposed to be the real part of
$\Phi$ and with minimal supergravity its mass-squared  vanishes
provided that \cite{holman,graham}
$\Phi_0=\Mpl$. Including the quartic 
term in $K$ the same thing will occur for some other value $\Phi_0=B$,
where $B$ depends on the coefficient of the quartic term but is still of 
order $\Mpl$. So $\eta$ will be accidentally
suppressed if $|\Phi_0-B|\ll \Mpl$.

Barring an accidental cancellation, the smallness of $\eta$
requires a non-generic form for the supergravity potential.
With minimal supergravity, a simple choice that works is
$W=V_0^{1/2}\Phi$, which leads to the global supersymmetry result,
corresponding to an exactly flat potential
\cite{CLLSW}, $V=V_0$.
Taking $W=V_0^{1/2}\Phi(1-\frac12\mu\Phi^p)$ could then give \cite{kumekawa}
a non-hybrid inflation 
potential $V\simeq V_0(1-\mu\phi^p)$. Alternatively,
taking $W$ of the form (\ref{wglobal}) 
could  give \cite{CLLSW} hybrid 
inflation with $\Phi_1=\Phi_2=0$, leading again to 
$W=V_0^{1/2}\Phi$ and an absolutely flat potential
whose slope might come from a loop 
correction \cite{qaisar,dvaliloop,qaisarlatest}.
But the quartic contribution to $K$ is expected to spoil minimal
supergravity, in which case these choices will not work.\footnote
{A possible intermediate 
strategy \cite{izawa} might be to use one of these forms to
eliminate the contribution to $\eta$ of the minimal (quadratic) term,
and then suppose that the quartic term is accidentally
suppressed. Such an accidental {\em suppression} is perhaps
less objectionable than the accidental {\em cancellation} which is 
otherwise required.
However, in the model of Ref.~\cite{izawa} the suppression
needs to be considerable, and the other parameters also have to be 
chosen delicately to get a viable model. If one were to apply the idea
to the model of Refs.~\cite{CLLSW,qaisar,dvaliloop},
with the slope coming from
a loop correction, the suppression would need to be severe since 
$\eta\sim .01$.} In the context of nonminimal supergravity
there are the following proposals, mostly involving things that have 
been mentioned earlier.
\begin{enumerate}
\item
The potential might be dominated by the 
$D$ term rather than by the $F$ term \cite{ewansgrav} in which case its
variation will probably be dominated by a loop correction 
\cite{bindvali,halyo} of the form
$A\ln(\phi/B)$, giving
$n=.96$ to $.98$.
\item
One can impose restrictions on the form of the $F$ term
\cite{ewansgrav}, which are of a type that might emerge
from superstring theory provided that it is in a perturbative regime.
This scheme is sufficiently flexible that it can accomodate
practically 
all versions of hybrid inflation \cite{ewansgrav,mutated,ournew}, as
well as the large kinetic term model of Ref.~\cite{ewansgrav}.
Thus it can accomodate more or less any measured value of $n$.
\item
It has been known for a long time that supergravity models
of the `no-scale'
type possess a `Heisenberg symmetry' that can eliminate the 
mass-squared of order $H^2$ for generic scalar fields. The problem has 
been to stabilize the `Polonyi' or (in the modern setting) modulus
field occurring in such theories. It was known from the beginning that
an ad hoc prescription can fix the modulus to give  viable models
of inflation, though the focus was not on small-field models
(for reviews see Refs.~\cite{olive,abook} and for a different ad hoc 
prescription see Ref.~\cite{hitoshi1}).
Recently it has been 
noted \cite{gaillard}
that a loop correction might provide the stabilization
automatically. As yet no model of inflation based on this latter scheme
has 
been proposed. Since the modulus will adjust itself to
the current minimum of the potential during inflation 
\cite{hitpers} 
(instead of being absolutely fixed as in the early ad hoc schemes)
one may be looking at something resembling
a mutated hybrid inflation model.
\item
The tree-level contribution to $\eta$ might be cancelled, within a 
limited interval of $\phi$, by a loop correction 
\cite{ewanloop},
leading to an inverted quadratic potential with
$n$ considerably below 1.
\item
Identifying the inflaton field with a pseudo-Goldstone boson 
keeps its potential absolutely flat in the limit of unbroken symmetry.
Explicitly breaking the symmetry 
\cite{natural,natural2,natural4,kinney,glw,ewanprep} then gives a 
nonzero $\eta$, which can be small if 
the symmetry is broken only by $W$ 
and not by $K$, or if most of the potential comes from another field
\cite{ewanprep}.
This solution to the problem would 
probably give $n$ significantly below 1 in the former  case, but 
indistinguishable from 1 in the latter.
\item
The form of the potential $V(\phi)$ 
might depend on some other field, which is driven
to a value corresponding to negligible $\eta$ before cosmological
inflation starts. This is the proposal of Ref.~\cite{grahamnew}
mentioned in Section IV, 
leading to 
$n=.84$ to $.92$. For the proposal to work 
one still needs a mechanism
for keeping the potential flat in the direction of the other field
(a global $U(1)$ in the above model)  so 
it might become the inflaton
in a different regime of parameter space.
\end{enumerate}

\subsection{Inflation as a probe of supergravity}

From these considerations we see that inflation is a very powerful probe 
of supergravity. In most models, the $F$ term 
dominates the potential, and $\eta$ generically receives
various contributions of 
order $1$ which must cancel. To keep it small requires 
either an accidental cancellation (reasonable only if $\eta$ is not 
too small) or non-generic 
forms for $K$ and $W$. In constructing suitable forms, 
it is not permissible to  make the `minimal supergravity' 
approximation of
ignoring the higher order terms in the expansion
(\ref{kahlerexp}) of $K$. 

Having ensured the smallness of $\eta$, one is not in general left with 
the global supergravity result $V=\sum|W_n|^2$. It
does indeed hold for solution 2 of the problem,
but both $W$ and $K$ need to have very special forms.

Finally, if 
the $D$ term dominates there is no problem about keeping $\eta$ small,
but to require this is itself a strong constraint
on the model.

\section{Inflation with a multi-component inflaton}

So far we have assumed that the slow-rolling inflaton field is 
essentially unique. What does `essentially' mean in this context?
A strictly unique inflaton trajectory would be one lying
in a steep-sided valley in field space. This is not very likely in a 
realistic model. Rather there will be a whole 
family of possible inflaton trajectories, lying in the space of two or 
more real fields $\phi_1$, $\phi_2, \cdots$.  
In this case, it quite useful to 
think of the inflaton field as a multi-component object with 
components $\phi_a$.
But even for a multi-component inflaton field it
may well be the case that different choices 
of trajectory lead to the same universe, and if that is the case
we still have an `essentially' unique inflaton field.

A familiar example of the essentially unique case
is if the
inflaton field
is the modulus of a complex field charged under a global $U(1)$
symmetry.
Then the possible inflaton trajectories are the radial
lines and the $U(1)$ symmetry ensures that all trajectories correspond
to identical universes during inflation. Moreover, the transition
to  a universe composed of matter plus radiation 
will be the same for all of them. Thus the inflaton field 
is essentially unique.

More generally, the inflaton field 
will be essentially unique if the 
inflaton trajectories are practically straight in the space of
canonically normalized fields, and if also the transition from
inflation to a matter/radiation universe is the same for each 
trajectory. 

The essentially non-unique case is when different possible
trajectories correspond to differently evolving universes.
Then trajectories near the classical one 
cannot be ignored,
because the quantum fluctuation
kicks the inflaton field onto them in a random way.
From now on, `multi-component' will refer to this case.

Even in the single-component case there is a quantum fluctuation,
which kicks the inflaton back 
and forth along the classical trajectory. This causes an adiabatic 
density perturbation, and in general the 
effect of the orthogonal fluctuation onto nearby trajectories is to 
cause an additional adiabatic density perturbation.
(Exceptionally it might cause an isocurvature perturbation as
mentioned at the end of this section.)

Multi-component inflaton models generally have just two components,
and are called double inflation models because the trajectory can
lie first in the direction of one field, then in the direction of the 
other.  They were first proposed in
the context of non-Einstein gravity 
\cite{starob85,d1,d2,d4,d5,d11,noncanon,davidjuan}.
By redefining the fields
and the spacetime metric one can recover Einstein gravity,
with fields that are not small on the Planck scale and
in general non-canonical kinetic terms and a non-polynomial 
potential. Then models 
with canonical kinetic terms were proposed 
\cite{c3,d7,d9,d10,d13,c1,c2,d3,d6,d8,isocurv,salopek95},
with potentials such as
$V=\lambda_1\phi_1^p+\lambda_2\phi_2^q$.
These potentials too inflate in the large-field regime
where theory provides no guidance about the form of the potential.
However there
seems to be no bar to having a small-field multi-component model,
and one may yet emerge in 
a well-motivated particle theory setting.
In that case a hybrid model might emerge, though 
the models proposed so far 
are all of the 
non-hybrid type (ie., the multi-component inflaton is entirely 
responsible for the potential).

In this brief survey we have focussed on the era when cosmological 
scales leave the horizon. In the hybrid inflation
model of Ref.~\cite{lisa,glw}, 
the `other' field is responsible for the last several $e$-folds
of inflation, so one is really dealing with a
two-component inflaton (in a non-hybrid model). 
The scales corresponding to the last 
few $e$-folds are many orders of magnitude shorter than
the cosmological scales, but it turns out that the perturbation
on them is big so that black holes can be produced.
This phenomenon was investigated in Refs.~\cite{lisa,glw}.
The second reference also investigated the possible production of
topological defects, when the first field is destabilized.

\subsection{Calculating the spectrum of the curvature perturbation}

We noted at the beginning of Section II that the adiabatic density 
perturbation on scales well outside the horizon 
is specified by a quantity $\cal R$, which 
defines the curvature of comoving hypersurfaces.
Going back in time from the epoch of horizon entry, it soon achieves a 
constant value, which is maintained 
at least until the beginning of the 
radiation-dominated era preceding the present matter-dominated 
one, and unless otherwise stated $\cal R$ denotes this constant value.
On the assumption that it is a gaussian random field, which is 
generally the case if it is generated by a vacuum fluctuation of the 
inflaton field, its stochastic properties are completely determined 
by its spectrum
${\cal P}_{\cal R}$. 

In this section we will see how to calculate the spectrum,
first for a single-component inflaton and then for a multi-component
one. In both cases we use an approach that has only recently been 
developed \cite{salopek95,ewanmisao}, though its starting point
can already be seen in the first derivations of the spectrum
\cite{hawking,starob82,guthpi}.
This is the assumption that
after smoothing on a scale well outside the 
horizon (Hubble distance) the evolution of the universe 
along each comoving worldline will be practically 
the same as in a Robertson-Walker
universe.\footnote
{`Smoothing' on a scale $R$ means that one replaces (say) $\rho({\bf x})$ 
by $\int d^3x' W(|{\bf x}'-{\bf x}|) \rho({\bf x}')$ with 
$W(y)\simeq 1$ for $y\lsim R$ and
$W\simeq 0$ for $y\gsim R$. A simple choice is to take
$W=1$ for $y<R$ and $W=0$ for $y>R$
(top-hat smoothing).}
One always makes such an assumption when doing
cosmological perturbation theory. When a quantity is split into an
average plus a perturbation, the average is
identified with the `background' quantity that is supposed to correspond 
to a Robertson-Walker  universe (ie., to an absolutely homogeneous and 
isotropic one). 
And it is accepted without question that 
the criterion for the size of the averaging region is that it be much 
bigger than the horizon.

The averaging scale will of course be chosen to be a comoving one.
When splitting a quantity into an average and a perturbation it is 
usually taken to be much bigger than the one corresponding to the 
whole presently observable universe. But
in the early universe it makes equal sense 
to make it much smaller.
This is indeed often done implicitly (sometimes explicitly
\cite{myaxion}) for a {\em single} small region, namely the one around 
us. The crucial idea behind the present approach 
\cite{salopek95,ewanmisao} is to recognize that the comparison of 
{\em different} regions provides a simple and powerful technique for
calculating the density perturbation. 

As we discuss later, it is quite different from the
usual one of writing down, and then solving, a closed set of equations
for the perturbations in the relevant degrees of freedom
(for instance the components of the inflaton field during inflation).
Roughly speaking the present approach replaces the sequence `perturb then solve'
by the far simpler sequence `solve then perturb', though it is actually 
more general than the other approach. 
For the case of a single-component inflaton it gives 
a very simple, and completely general, proof of 
the constancy of $\cal R$ on scales well outside the horizon.
For the multi-component case it allows one to follow the evolution of
$\cal R$, knowing only the evolution of the {\em unperturbed}
universe corresponding to a given value of the initial inflaton field.
So far
it has been applied to three
multi-component models \cite{salopek95,davidjuan,glw}.

\subsection{The case of a single-component inflaton}

We begin with a derivation of the usual result for the single-component 
case. The assumption about the evolution along each comoving worldline
is invoked only at the very end,
when it is used to establish the constancy of $\cal R$
which up till now has only been demonstrated for special cases.
Otherwise the proof is the standard one \cite{LL2}, but it provides a 
useful starting point for the multi-component case.

A few Hubble times 
after horizon exit during inflation, when $\cal R$ 
can first be regarded as a classical quantity, its spectrum 
can be calculated using the relation
\cite{kodamasasaki,LL2}\footnote
{In \cite{LL2} there is an incorrect minus sign on the right hand side.}
\be
{\cal R}({\bf x})=H\Delta\tau({\bf x})
\label{r1}
\ee 
where $\Delta\tau$ is the separation of the comoving hypersurface
(with curvature $\cal R$)
from a spatially flat one coinciding with it on average.
The relation is generally true, but we apply it at an epoch
a few Hubble times after horizon exit during inflation.

On a comoving hypersurface the inflaton field $\phi$
is uniform, because the momentum density $\dot\phi {\bf \nabla}\phi$
vanishes. It follows that
\be
\Delta\tau({\bf x})=-\delta\phi({\bf x})/\dot\phi
\ee
where now $\delta\phi$ is defined on the  flat hypersurface.
Note that the comoving hypersurfaces become singular (infinitely
distorted) in the slow-roll limit $\dot\phi\to 0$, so that 
to first order in slow-roll any non-singular choice of hypersurface 
could actually be used to define $\delta\phi$.

The spectrum of $\delta\phi$ 
is calculated by assuming that well before horizon entry
(when the particle concept makes sense) $\delta\phi$ is a 
practically free field in the vacuum state. Using the flatness and
slow-roll conditions one finds, a few Hubble times after horizon 
exit, the famous result \cite{LL2}
${\cal P}_\phi=(H/2\pi)^2$, which leads to the 
usual formula (\ref{delh}) for the spectrum.

However, this result refers to $\cal R$ a few Hubble times after horizon
exit, and we need to check that $\cal R$ remains constant until
the radiation dominated era where we need it. To  calculate the rate of 
change of $\cal R$ we proceed as follows \cite{lythmuk,LL2}.

In addition to the energy density $\rho$ and 
the pressure $P$, we consider a locally defined
Hubble parameter 
$H=\frac13D_\mu u^\mu$ where $u^\mu$ is the four-velocity of 
comoving
worldlines and $D_\mu$ is the covariant derivative.
Our $3H$ is often called $\theta$ in the literature.
The universe is sliced into comoving hypersurfaces, and each 
quantity is split into an average (`background') plus a 
perturbation, $\rho({\bf x},t)=\rho+\delta\rho({\bf x},t)$
and so on. (We use the same symbol for the local and the background 
quantity since there is no confusion in practice.)
As usual, $\bf x$ is the Cartesian position-vector of a
comoving worldline and $t$ is the time. As we are working to first order 
in the perturbations they `live' in unperturbed spacetime.

The locally defined quantities satisfy 
\cite{hawkingellis,lyth85,lythmuk,LL2}
\be
H^2=\Mpl^{-2}\rho/3-\frac23\frac{k^2}{a^2}\cal R
\label{localfr}
\ee
(The equation is valid for each Fourier mode, and also for the
full quantities with the identification
$(k/a)^2=-\nabla^2$. From now on that is taken for granted.)
This is the Friedmann equation except that 
$K\equiv(2/3)k^2\cal R$ need not be constant.
The evolution along each worldline is
\bea
\frac{d\rho}{d\tau}&=&-3H(\rho+P)
\label{cont}\\
\frac{d H}{d\tau} &=& -
H^2 - \frac12 \Mpl^{-2} (\rho+3P)
+\frac13\frac{(k/a)^2\delta P}{\rho+P}
\label{raych}
\eea
Except for the last term these are the same as in an unperturbed 
universe. If it vanishes $\cal R$ is constant, but otherwise one finds
\be
\dot{\cal R}=-H\delta P/(\rho +P)
\label{rdot}
\ee
In this equation we have in mind that $\rho$ and $P$ are background quantities,
though as we are working to first order in the perturbations it would 
make no difference if they were the locally defined quantities.

The equation shows that $\cal R$ will be constant if $\delta P$ is negligible.
We now show that this is so, by first demonstrating that 
$\delta\rho$ is negligible, and then using the new viewpoint
to see that $P$ will be a practically 
unique function of $\rho$ making $\delta P$ also negligible.

Extracting the perturbations from
Eq.~(\ref{localfr}) gives 
\be
2\frac{\delta H}{H}=\frac{\delta\rho}{\rho } -\frac23\left(
\frac{k}{aH}\right)^2 \cal R
\label{perturbfre}
\ee
This allows one to calculate the evolution of $\delta \rho$ from
Eq.~(\ref{cont}), but we have 
to remember that the proper-time separation of the hypersurfaces
is position-dependent. Writing
$\tau({\bf x},t)=t+\delta\tau({\bf x},t)$ we have
\cite{kodamasasaki,lythmuk,lyst} 
\be
\delta(\dot\tau)= -\delta P/(\rho+P)
\label{deltataudot}
\ee
Writing $\delta\rho/\rho\equiv (k/aH)^2 Z$ 
one finds \cite{lythmuk}
\be
(fZ)'=f (1+w) \cal R
\label{zexpr}
\ee
Here a prime denotes $d/d(\ln a)$ and 
$f'/f\equiv(5+3w)/2$ where $w\equiv P/\rho$.
With $w$ and $\cal R$ constant, and dropping a decaying mode, this gives
\be
Z=\frac{2+2w}{5+3w} \cal R
\ee
More generally, integrating Eq.~(\ref{zexpr})
will give $|Z|\sim |\cal R|$ for any reasonable variation of
$w$ and $\cal R$. Even for a bizarre variation there is no scale 
dependence in either $w$ (obviously) or in $\cal R$ (because 
Eq.~(\ref{penult}) gives it in terms of $\delta P$, and we will see that
if $\delta P$ is significant it is scale-independent).
In all cases
$\delta\rho/\rho$ becomes negligible on 
scales sufficiently far outside the horizon.

The discussion so far applies to each Fourier mode separately, on 
the assumption that the corresponding perturbation is small.
To make the final step, of showing that $\delta P$ is also negligible,
we need to consider the full quantities $\rho({\bf x},t)$ 
and so on. But we still want to consider only scales that are well
outside the horizon, so we suppose that all quantities are smoothed
on a comoving scale somewhat shorter than the one of interest.
The smoothing removes Fourier modes on scales shorter than the 
smoothing scale, but has practically no effect on the scale
of interest. 

Having done this, we invoke the assumption that the evolution of the 
universe along each worldline is practically the same as in an unperturbed 
universe. In the context of slow-roll inflation, this means that 
the evolution is determined by the  inflaton field
at the `initial' epoch a few Hubble times after horizon exit.
To high accuracy, $\rho$ and $P$  are well
defined functions of the initial inflaton field and
{\em if it has only one component}
this means that they are well defined 
functions of each other. Therefore $\delta P$ will be very small
on comoving hypersurfaces because $\delta \rho$ is.\footnote
{If $k/a$ is the smoothing scale, the 
assumption that the evolution is the same as in an unperturbed
universe with the same initial inflaton field
has in general errors of order $(k/aH)^2$. 
In the single-component case, where $\delta P$ is also of this order,
we cannot use the assumption to actually 
calculate it, but neither is it of any interest.}

Finally, we note for future reference that $\delta H$ is also
negligible because of Eq.~(\ref{perturbfre}).

\subsection{The multi-component case}

It is assumed that while cosmological scales are leaving the horizon
all components of the inflaton have the slow-roll behaviour
\be
3H\dot \phi_a = - V_{,a}
\ee
(The subscript $,a$ denotes the derivative with respect to $\phi_a$.)
Differentiating this
and comparing it with the exact expression $\ddot\phi_a
+3H\dot \phi_a +V_{,a}=0$ gives consistency provided that
\bea
\Mpl^2 (V_{,a}/V)^2 &\ll& 1\\
\Mpl^2 |V_{,ab}/V| &\ll& 1
\label{flat2}
\eea
(The second condition could actually be replaced by a weaker one but let 
us retain it for simplicity.)
One expects slow-roll to hold if these flatness conditions are 
satisfied. Slow-roll plus the first flatness condition
imply
that $H$ (and therefore $\rho$) is slowly varying, giving 
quasi-exponential inflation. The second flatness condition 
ensures that $\dot\phi_a$ is slowly varying.

It is not necessary to assume that
all of the fields continue to slow-roll after cosmological scales 
leave the horizon. For 
instance, one or more of the fields might start to oscillate, while the others
continue to support quasi-exponential inflation, which ends only
when slow-roll fails for all of them.
Alternatively, the oscillation of 
some field might briefly interrupt inflation, which resumes when its 
amplitude becomes small enough. (Of course these things might happen 
while cosmological scales leave the horizon too, but that case 
will not be considered.)

The expression (\ref{r1}) for 
$\cal R$ still holds in the multi-component case. Also, 
one still has 
$\Delta\tau=-\delta\phi/\dot\phi$ if $\delta\phi$ 
denotes the component of the vector $\delta\phi_a$
parallel to the trajectory. (The momentum density 
seen by an observer orthogonal to an arbitrary hypersurface is
$\dot\phi_a {\bf \nabla} \phi_a$.) A few Hubble times after horizon exit
the spectrum of 
every inflaton field component, in particular the parallel one,
is still $(H/2\pi)^2$. If $\cal R$ had no subsequent variation this 
would lead to the usual prediction, but we are considering the case
where the variation is significant.
It is given in terms of $\delta P$ by 
Eq.~(\ref{rdot}), and when $\delta P$ is significant it can be calculated
from the assumption that the evolution along each
worldline is the same as for an unperturbed universe with the same
initial inflaton field. This will give 
\be
\delta P=P_{,a}\delta\phi_a
\ee
where $\delta\phi_a$ is evaluated at the initial epoch and the 
function $P(\phi_1,\phi_2,\cdots,t)$ represents the evolution of 
$P$ in an unperturbed universe. Choosing the basis so that one of 
the components is the parallel one, and remembering that all components
have spectrum $(H/2\pi)^2$, one can calculate 
the final spectrum of $\cal R$. The only input is the evolution of 
$P$ in the unperturbed universe corresponding to a generic initial 
inflaton field (close to the classical initial field).

In this discussion we started with Eq.~(\ref{r1}) for the initial
$\cal R$, and then invoked Eq.~(\ref{rdot}) to evolve it.
The equations 
can actually be combined to give 
\be
{\cal R}=\delta N
\ee
where $N=\int Hd\tau$ is the number of Hubble times 
between the initial flat hypersurface and the final comoving one on 
which $\cal R$ is evaluated. This remarkable expression was given
in Ref.~\cite{starob85} and proved in Refs.~\cite{salopek95,ewanmisao}.
The approach we are using is close to the one in the
last reference.

The proof that
Eqs.~(\ref{r1}) and (\ref{rdot}) lead to ${\cal R}=\delta N$ 
is very simple. First
combine them to give
\be
{\cal R}({\bf x},t)=H_1\Delta\tau_1({\bf x})-
\int_{t_1}^{t} H(t) \frac{\delta P}{\rho+P}
\label{penult}
\ee
where $t_1$ is a few Hubble times after horizon exit.
Then use Eq.~(\ref{deltataudot}) to give
\be
{\cal R}({\bf x},t)=H_1\Delta\tau_1({\bf x})+
\int_{t_1}^{t} H(t) \delta \dot\tau({\bf x},t)dt
\label{penult1}
\ee
Next note that because $\delta H$ is negligible
this can be written
\be
{\cal R}({\bf x},t)=H_1\Delta\tau_1({\bf x})+
\delta \int_{t_1}^{t} H({\bf x},t) \dot\tau({\bf x},t)dt
\label{ult}
\ee
Finally redefine $\tau({\bf x},t)$ so that it
vanishes on the
initial {\em flat} hypersurface, which gives the desired 
relation ${\cal R} = \delta N$.

In Ref.~\cite{ewanmisao} this relation is derived using an 
arbitrary smooth interpolation of hypersurfaces
between the initial and final one, rather than by making the sudden jump 
to a comoving one. Then $H$ is replaced by the corresponding quantity 
$\tilde H$ 
for worldlines orthogonal to 
the interpolation (incidentally making $\delta \tilde H$ non-negligible).
One then finds $\cal R=\delta\tilde N$. One also finds that the right 
hand side is independent of the choice of the interpolation, as it must 
be for consistency. If the interpolating hypersurfaces are chosen to be
comoving except very near the initial one, $\tilde N\simeq N$ which 
gives the desired formula ${\cal R}=\delta N$.\footnote
{The last step is not spelled out in Ref.~\cite{ewanmisao}.
The statement that $\tilde N$ is independent of the interpolation 
is true only on scales well outside the 
horizon, and its physical interpretation is unclear though it drops out 
very simply in the explicit calculation.}

\subsection{Calculating the spectrum and the spectral index}

Now we derive explicit formulas for the spectrum and the spectral index, 
following \cite{ewanmisao}. 
Since the evolution of 
$H$ along a 
comoving worldline will be the same as for a homogeneous universe
with the same initial inflaton field, $N$ is a function only of this
field and we have
\be
{\cal R}=N_{,a} \delta\phi_a  
\ee
(Repeated indices are summed over
and the subscript $,a$ denotes differentiation with respect to $\phi_a$.)
The perturbations $\delta\phi_a$ are Gaussian random fields generated by the
vacuum fluctuation, and have a common 
spectrum $(H/2\pi)^2$. The
spectrum $\delta_H^2\equiv (4/25){\cal P}_{\cal R}$
is therefore 
\be
\delta_H^2= \frac{V}{75\pi^2\Mpl^2}N_{,a}N_{,a}
\ee

In the single-component case, $N'=\Mpl^{-2} V/V'$
and we recover the usual expression. In the multi-component case
we can always choose the basis 
fields so that while cosmological scales are leaving the horizon
one of them points along the inflaton trajectory, and then its
contribution gives the standard result with the orthogonal
directions giving an additional contribution.
Since the spectrum of gravitational waves is independent of the number 
components (being equal to a numerical constant times $V$)
the relative contribution $r$ of gravitational waves to the cmb
is always {\em smaller} in the multi-component case.

The contribution from the orthogonal directions
depends on the whole inflationary potential after the relevant scale
leaves the horizon, and maybe even on the evolution of the energy 
density after inflation as well. This is in contrast to the contribution 
from the parallel direction which depends 
only on $V$ and $V'$ evaluated
when the relevant scale leaves the horizon.
The contribution from the orthogonal directions will be at most of order
the one from the parallel direction provided that all $N_{,a}$ are at 
most of order $\Mpl^{-2} V/V'$.
We shall see later that this is a 
reasonable expectation at least if $\cal R$ stops varying after the end 
of slow-roll inflation.

To calculate the spectral index we need the analogue of
Eqs.~(\ref{nexpression}) and (\ref{Nrelation}). Using
the chain rule and $dN=-Hdt$ one finds
\bea
\frac{d}{d\ln k}
&=&-\frac{\Mpl^2}{V}V_{,a} \frac{\partial}{\partial\phi_a}\\
N_{,a} V_{,a} &= & \Mpl^{-2} V
\label{nv}
\eea
Differentiating the second expression gives
\be
V_{,a} N_{,ab}+ N_{,a} V_{,ab} = \Mpl^{-2}V_{,b}
\ee
Using these results one finds
\be
n-1 = -\frac{\Mpl^2V_{,a}V_{,a}}{V^2}
-\frac2{\Mpl^2N_{,a}N_{,a}}
+2\frac{\Mpl^2N_{,a} N_{,b} V_{,ab} }
{VN_{,d}N_{,d}}
\label{multin}
\ee
Again, we recover the single field case using $N'=\Mpl^{-2} V/V'$.

Differentiating this expression and setting $\Mpl=1$ for clarity
gives
\bea
\frac{dn}{d\ln k}&=&
-\frac {2}{V^3}V_{,a} V_{,b} V_{,ab}
+\frac2{V^4} (V_{,a}V_{,a})^2
+\frac 4 V \frac{(V-N_{,a}N_{,b} V_{,ab} )^2}{(N_{,d}N_{,d})^2}\nonumber\\
&+&\frac2 V \frac{N_{,a} N_{,b} V_{,c} V_{,abc} }{N_{,d}N_{,d}}
+\frac4 V \frac {( V_{,c} -N_{,a} V_{,ac}) N_{,b} V_{,bc} }
{N_{,d}N_{,d}}
\eea
A correction
to the formula for $n-1$ has also been worked out \cite{nak}. 
Analogously with the single-component case, both this correction 
and the variation of $n-1$ involve the first, second and third 
derivatives of $V$.
Provided that the derivatives of $N$ in the orthogonal directions are 
not particularly big, and barring cancellations, 
a third flatness 
condition 
$V_{,abc}V_{,c}/V^2\ll \max\{\sum_a(V_{,a})^2,\sum_{ab}|V_{,ab}|\}$
ensures that both the correction
and the variation of $n-1$ in  a Hubble time are small.
(One could find a weaker condition that would do the same job.)

These formulas give the spectrum and spectral index of the 
density perturbation, if one knows the evolution
of the homogeneous universe corresponding both to the
classical inflaton trajectory and to nearby trajectories.
An important difference in principle from the single-component
case, is that the classical trajectory is not uniquely specified by 
the potential, but rather has to be given as a separate piece of 
information. However, if there are only two components the classical 
trajectory can be determined from the COBE normalization of the 
spectrum, and then there is still a prediction for the spectral index.

This treatment 
can be generalized straightforwardly \cite{ewanmisao}
to the case of
non-canonical kinetic terms of the form (\ref{hmetric}), that is 
expected in 
supergravity.
However, in the small-field regime one expects the 
curvature associated
with the `metric' $h_{ab}$ to be negligible, and then one can recover 
the canonical normalization 
$h_{ab}=\delta_{ab}$ by redefining the 
fields.

\subsection{When will $\cal R$ become constant?}

We need to evaluate $N$ up to the epoch where ${\cal R}=\delta N$ 
has no further time dependence. When will that be?

As long as all fields are slow-rolling, $\cal R$ is constant if and only 
if the inflaton trajectory is straight.
If it turns through a small angle $\theta$,
and the trajectories have not converged appreciably since horizon exit,
the fractional change in $\cal R$ is in fact $2\theta$.\footnote
{Thinking in two 
dimensions and taking the trajectory to be an arc of a circle,
a displacement  $\delta\phi$
towards the center decreases 
the length of the trajectory by an amount $\theta\delta\phi$,
to be compared with the decrease $\delta\phi$ for the same
displacement along the trajectory. (The rms displacements will indeed
be the same if the trajectories have not converged.) Since 
The 
speed along the new trajectory is faster in inverse proportion to the 
length since it is proportional to $V'$
and $V$ is fixed at the 
initial and final points on the trajectory. Thus the perpendicular 
displacement increases $N$ by $2\theta$ times the effect of a 
parallel displacement, for $\theta\ll 1$.}
Since slow-roll
requires that the change in the vector $\dot\phi_a$ during 
one Hubble time is negligible, the total angle turned is $\ll N$.
Hence the relative contribution of the orthogonal directions cannot
be orders of magnitude bigger than the one from the parallel 
direction, if it is generated during slow-roll inflation.
(In two dimensions the angle turned cannot exceed $2\pi$ of course, but 
there could be say a corkscrew motion in more dimensions.) 
Later slow-roll may fail for one or more of the fields,
with or without interrupting inflation, and things become more 
complicated, but in general there is no reason why $\cal R$ should stop varying
before the end of inflation.

Now let us ask what happens after the end of inflation 
(or to be more precise,
after significant particle production has spoiled the above analysis,
which may happen a little before the end).
The simplest case is if 
the relevant trajectories have practically converged to a 
single trajectory $\phi_a(\tau)$, as in Ref.~\cite{glw}.
Then $\cal R$ will not vary any more (even after inflation is over)
as soon as the trajectory has
been reached. Indeed, setting $\tau=0$ at the end of 
inflation, this unique trajectory corresponds to a post-inflationary universe 
depending only on $\tau$. The fluctuation in the initial field values 
causes a fluctuation $\Delta\tau$ 
in the arrival time at the end of inflation, leading to a 
time-independent ${\cal R}=\delta N
=H_{\rm end}\Delta\tau$. 

What if the trajectory is not unique at the end of inflation?
After the completion of the transition from inflation to a universe of 
radiation and matter, Eq.~(\ref{rdot}) tells us that
$\cal R$ will be constant provided that $\delta P$ and $\delta \rho$
are related in a definite way. This is the case 
during matter domination
($\delta P=0$) or radiation domination 
($\delta P=\frac13\delta\rho$).\footnote
{If the `matter' consists of a nearly homogeneous oscillating
scalar field
one actually has $\delta P=\delta\rho$ (=$\frac12
\delta (d\phi/d\tau)^2$), since on comoving hypersurfaces
$\phi$ and therefore $V$ is constant. But
this still makes $\delta P$ negligible.}

Immediately following inflation there might be 
a quite complicated situation, with `preheating' \cite{kls}
or else the quantum fluctuation of the `other' field in hybrid models
\cite{CLLSW} converting most of the inflationary potential energy
into marginally relativistic particles in much less than a Hubble time.
But after at most a few Hubble times
one expects to 
arrive at a matter-dominated era so that $\cal R$ is constant.
Subsequent events will not cause $\cal R$ to vary provided that they 
occur at definite values of the energy density, since again $P$ will 
have a definite relation with $\rho$.
This is indeed the case for the usually-considered events, such as 
the decay of
matter into radiation and thermal
phase transitions
(including thermal inflation).
The conclusion is that it is 
reasonable to suppose that $\cal R$ achieves a constant
value at most a few Hubble times after inflation.
On the other hand one cannot exclude the possibility that one of the 
orthogonal components of the inflaton provides a significant additional 
degree of freedom, allowing $\cal R$ to have additional variation before 
we finally arrive at the radiation-dominated era preceding the present 
matter-dominated era. 

This leaves the transition, lasting probably at most a few Hubble times,
from inflation to the first epoch of matter domination.
It is conjecture in Ref.~\cite{nak} that
the variation of $\cal R$ during the transition
may still be negligible
if slow-roll holds almost until the end of inflation.
(This is in the present context of the first-order slow-roll 
calculation, not the second-order one that is the main focus
of Ref.~\cite{nak}.) 
The question requires detailed study however.

\subsection{Working out the perturbation generated by slow-roll
inflation}

If $\cal R$ stops varying by the end of inflation, the
final hypersurface can be located
just before the end (not necessarily at the very end
because that might not correspond to a hypersurface of constant energy
density). Then, knowing the potential
and the hypersurface in field space that corresponds to the end of
inflation, one can work out $N(\phi_1,\phi_2,\cdots)$
using the equations of motion for the fields, and the expression
\be
3\Mpl^2 H = \rho =V+\frac12\frac{d\phi_a}{d\tau}\frac{d\phi_a}{d\tau}
\ee

To perform such a calculation it is not necessary that
all of the fields continue to slow-roll after cosmological scales leave
the horizon. In particular, the oscillation of
some field might briefly interrupt inflation, which resumes when its
amplitude becomes small enough.
If that happens it may be necessary to take
into account `preheating' during the interruption.

In general all this is quite complicated, but there is one 
case that may be extremely simple, at least 
in a limited regime of parameter space.
This is the case
\be
V=V_1(\phi_1) + V_2(\phi_2) + \cdots
\ee
with each $V_a$ proportional to a power of $\phi_a$.
For a single-component inflaton this gives inflation ending at 
$\phi_{\rm end} \simeq \Mpl$, with cosmological scales leaving
the horizon at $\phi\gg\phi_{\rm end}$.
If the potentials $V_a$ are identical we recover that case.
If they are different, slow-roll
may fail in sequence for the different components, 
but in some regime of parameter space
the result for
$N$ (at least) might be
the same as if it failed simultaneously for all components.
If that is the case one can derive simple formulas \cite{d7,salopek95},
provided that cosmological scales leave the horizon
at $\phi_a\gg \phi_a^{\rm end}$ for all components.

One has
\be
Hdt=-\Mpl^{-2}\frac{V}{V_1'}d\phi_1=
-\Mpl^{-2}\sum_a \frac{V_a}{V_a'} d\phi_a
\ee
It follows that 
\be
N=\Mpl^{-2}\sum_a \int_{\phi_a^{\rm end}}^{\phi_a}\frac{V_a}{V_a'} d\phi_a
\ee
Since each integral is dominated by the endpoint $\phi_a$, we have 
$N_{,a} = \Mpl^{-2}V_a/ V_a'$ and
\be
\delta_H^2 =\frac{V}{75\pi^2\Mpl^6} \sum_a \left(\frac{V_a}{V_a'}
\right)^2
\ee
The spectral index is given by
Eq.~(\ref{multin}),
which simplifies slightly
because $V_{,ab}=\delta_{ab}V_a''$.

The simplest case is 
$V=\frac12m_1^2\phi_1^2+\frac12m_2^2\phi_2^2$.
Then $n$ is given by the following formula
\be
1-n= \frac1 N\left[
\frac{(1+r)(1+\mu^2 r)}{(1+\mu r)^2} + 1\right]
\ee
where $r=\phi_2^2/\phi_1^2$ and $\mu=m_2^2/m_1^2$.
If $\mu=1$ this reduces to the single-component formula
$1-n=2/N$.
Otherwise it
can be 
much bigger, but note that our assumptions will be valid
if at all in a restricted region of the $r$-$\mu$ plane.

\subsection{Comparison with the usual approach}

The usual approach is to use cosmological perturbation theory to find
a closed system 
of linear equations, for perturbations in the relevant degrees of 
freedom. For each Fourier mode there is a set of coupled differential 
equations, which can 
be solved with enough effort. This approach
has been used to establish the constancy of $\cal R$ in special cases 
for a single-component inflaton \cite{usual}, and
to calculate $\cal R$ at the end of inflation with a multi-component 
inflaton \cite{c1,c2,c3,noncanon,davidjuan}. 
When it can be formulated, the usual approach
will lead to the same result as the present one, since
on scales far outside the horizon all spatial gradients in the equations
will become negligible.

On the other hand it is not clear that the desired closed system of 
equations will always exist.
It might for instance happen
that the small-scale quantum fluctuation of the `other' field
in a hybrid model
generates field gradients which play a crucial role in the transition 
from inflation to a matter-dominated universe \cite{CLLSW}.
In that case the evolution of (say) $P$ on large scales is sensitive
to the small-scale behaviour, and one will not be able to develop a 
closed system of equations for a given large-scale
Fourier component.

A striking demonstration of the greater 
power of the present approach
is provided by
the issue of the constancy
of $\cal R$ for a single-component inflaton. In the usual approach
one has to make simplifying assumptions, and even then 
the equations are 
sufficiently complicated that it is possible to make mistakes and
arrive at the incorrect 
conclusion that $\cal R$ is not constant \cite{grishchuk}. 
The present approach makes redundant all proofs of the constancy of 
$\cal R$ based on the usual approach.

\subsection{An isocurvature density perturbation?}

Following the astrophysics usage, we classify a density perturbation as 
adiabatic or isocurvature with reference to its properties well before 
horizon entry, but during the radiation-dominated era preceding the present 
matter-dominated era.\footnote
{In some of the theoretical literature this
kind of classification is made also at earlier times, in particular
during inflation, and also during the present matter-dominated era. 
From that viewpoint the density perturbation always starts out with
an isocurvature component in the multi-component case,
and it always ends up as adiabatic
by the present matter-dominated era.}
For an adiabatic density perturbation, the density of each particle 
species is a unique function of the total energy density. 
For an 
isocurvature density perturbation the total
density perturbation vanishes, but those of the individual particle 
species do not. 
The most general density perturbation is the sum of an adiabatic and
an isocurvature perturbation, with $\cal R$ specifying the adiabatic density 
perturbation only.

For an isocurvature perturbation to exist the universe 
has to possess more than the single degree of freedom provided by the 
total energy density.
If the inflaton trajectory is unique, or has become so by the end of 
inflation, there is only the single degree of freedom corresponding to 
the fluctuation back and forth along the trajectory and 
there can be no isocurvature perturbation. 
Otherwise one of the orthogonal fields
can provide the necessary degree of freedom.
The simplest way for this to happen is for the orthogonal field 
to survive, and acquire a potential so that it
starts to oscillate and becomes matter.\footnote
{If the potential  of the `orthogonal' field
already exists during inflation the inflaton
trajectory will have a tiny component in its direction, so that it is 
not strictly orthogonal to the inflaton trajectory. This makes no 
practical difference. In the axion case the potential is usually 
supposed to be generated by QCD effects long after inflation.}
The start of the oscillation 
will be determined by the total energy density, but its amplitude
will depend on the initial field value so there will be an isocurvature 
perturbation in the axion density. It will be compensated,
for given energy density, by the perturbations in the other species of 
matter and radiation which will continue to satisfy the adiabatic
condition $\delta\rho_m/\rho_m=\frac34\delta\rho_r/\rho_r$.

The classic example of this is the axion field
\cite{abook,kt,myaxion}, which is simple because
the fluctuation in the direction of the axion 
field causes no adiabatic density 
perturbation, at least in the models
proposed so far. The more general case, where one of the components of 
the inflaton may cause both an adiabatic and an 
isocurvature perturbation 
has been looked at in for instance Ref.~\cite{isocurv}, though not in the 
context of specific particle physics.
If an isocurvature perturbation in the non-baryonic dark matter density
is generated during inflation,
it must not conflict with observation and this imposes strong 
constraints on, for instance, models of the axion
\cite{myaxion,LIN2SC}. 

An isocurvature perturbation in the density of one a species of matter
may be defined by 
the `entropy perturbation' \cite{kodamasasaki,lyst,LL2}
\be
S= \frac{\delta\rho_m}{\rho_m}-\frac34\frac{\delta\rho_r}{\rho_r}
\ee
where $\rho_m$ is the non-baryonic dark matter density.
Equivalently, $S=\delta y/y$, where $y=\rho_m/\rho_r^{3/4}$.
Since we are dealing with scales far outside the horizon,
$\rho_m$ and $\rho_r$ evolve as they would in an unperturbed 
universe which means that $y$ is constant and so is $S$.
Provided that the field fluctuation is small $S$ will be proportional to 
it, and so will be a Gaussian random field with a nearly flat spectrum
\cite{myaxion,LL2}.

For an isocurvature perturbation, $\cal R$ vanishes 
during the radiation dominated era
preceding the present matter
dominated era. But on the
very large scales entering the horizon well after matter 
domination, $S$ generates a nonzero $\cal R$ during 
matter domination, namely
${\cal R}=\frac13 S$. A simple way of seeing this, which 
has not been noted before, is through the relation
(\ref{rdot}).
Since $\delta\rho=0$, one has $S=-(\rho_m^{-1}+\frac34\rho_r^{-1})
\delta\rho_r$. Then, using $\delta P=\delta\rho_r/3$, 
$\rho_r/\rho_m\propto a$ and $Hdt=da/a$ one finds
the quoted result by integrating Eq.~(\ref{rdot}).

As discussed for instance in Ref.~\cite{LL2}, the large-scale
cmb anisotropy coming from an isocurvature
perturbation is $\Delta T/T=-(\frac13+\frac 1{15})S$, where 
$S$ is evaluated on the last-scattering surface. The 
second 
term is the Sachs-Wolfe effect coming from the curvature perturbation we 
just calculated, and the first term is the anisotropy $\frac14
\delta\rho_r/\rho_r$ just after last scattering (on a comoving 
hypersurface). By contrast the anisotropy from an adiabatic perturbation
comes only from the Sachs-Wolfe effect, so for a given 
large-scale density 
perturbation the 
isocurvature perturbation gives an anisotropy six times bigger.
As a result
an isocurvature perturbation cannot be the dominant contribution to
the cmb, though one could contemplate a small contribution
\cite{iso}.

\section{Discussion and conclusion}

Let us summarize. Except in the last section we have focussed on models 
involving a single-component inflaton field, since they are simple
and give a relatively clean prediction for $n$.
The models considered include 
non-hybrid ones where the inflaton 
field $\phi$ dominates the potential during inflation, and hybrid ones
where this role is played by a different field $\psi$.
In the latter case $\psi$ minimizes the potential during inflation,
so that one still ends up with an effective potential $V(\phi)$.
It is usually assumed that
$V(\phi)$ (non-hybrid) or $V(\phi,\psi)$ (hybrid)
consists of one or a few low-order terms in a power-series expansion,
but this 
has motivation in the context of supergravity only if the fields are at 
most of order $\Mpl$. In the hybrid case one can still obtain
a non-polynomial $V(\phi)$ during inflation since $\psi$ may be a 
function of $\phi$ (the case of mutated hybrid inflation).
A non-polynomial potential can also emerge in a theory starting
out with non-canonical kinetic terms or non-Einstein gravity,
as well as from a quantum correction.

A mathematically simple potential giving inflation 
is the one first proposed to implement chaotic initial conditions
at the Planck scale, $V\propto \phi^p$, but with this potential
cosmological scales leave the horizon when
$\phi\sim 10\Mpl$. It
gives  $n-1=-(2+p)/(2N)$. It also gives  a significant
gravitational wave contribution to the cmb anisotropy, being practically
the only viable
potential proposed so far that does. (Extended inflation, which 
also gives a significant contribution,
is ruled out by observation
except in contrived versions.)

Virtually all other models so far proposed 
give, during inflation, a potential effectively 
of the form
$V=V_0(1\pm \mu\phi^p)$ with the constant term dominating.
With this potential the fields can be $\lsim \Mpl$, and in many
cases $\ll \Mpl$.
For the plus sign $p$ is a positive integer, but for the minus sign
it can be negative, or even non-integral.
In the last case, $p$ just below zero mimics the case of the potential
$V_0[1-A\ln(B/\phi)]$ and $p\to-\infty$ mimics
$V_0(1-e^{-q\phi})$, both of which
have good particle physics motivation (respectively from a
quantum correction or a theory starting out with non-canonical 
kinetic terms). 

Except for $0\lsim p\lsim 2$
the prediction for 
$n$ depends only on the exponent, as is shown in 
the Table for some integer values. The only important exceptions are the 
quadratic potentials
$V=V_0\pm \frac12m^2\phi^2$ which give
$n=1\pm 2\Mpl^2 m^2/V_0$. The flatness conditions 
require $\Mpl^2 m^2/V_0$ to be considerably less than 1.
Unless there is a reason why the parameters should be on the edge of 
the allowed region, one therefore 
expects $n$ to be indistinguishable from 1 in this case.
In the context of supergravity there may or may not be 
such a reason, depending on the model.

Looking at the Table, the
most striking thing is that most of the potentials make 
$n$ close to 1, but not very 
close. In fact practically all of the listed potentials are ruled out
unless $n$ lies in one of the intervals $.84<n<.98$ or $1.04<n<1.16$.
Another interesting feature is the dividing line 
between positive and negative values of $p$, which occurs 
somewhere in the range
$n=.92$ and $n=.96$.

Although any selection is as yet tentative, there is no doubt that some 
of these potentials are more favoured than others. 
Values of $n$ significantly bigger than 1 
seem unlikely; the potentials $V=V_0(1+\mu\phi^p)$ with $p>2$ are not 
favoured by particle theory, whereas the quadratic potential
is likely to give $n$ indistinguishable from 1.
 Values $n<.84$ are also unlikely, unless the 
inverted quadratic potential emerges from one of the 
non-hybrid settings discussed in Section IV.

This leaves the regime $.84<n\leq 1.00$, and within that are a few 
potentials
that might be regarded as
favoured theoretically. A very subjective
selection corresponds 
to the five rows marked of the Table marked by **.
The first and last cases are the quadratic and inverted quadratic 
potentials $V_0\pm\frac12m^2\phi^2$, thought of as being 
derived respectively
from 
ordinary and inverted hybrid inflation; since there is no reason
for the parameters to be on the edge of the region allowed by the 
flatness condition $\eta\ll 1$ one expects 
$n$ to be 
indistinguishable from 1 in these models. The second case is the 
loop-corrected potential that might arise if the $D$ term 
dominates, mimicked by the potential $V_0(1-\mu\phi^{-p})$
with $p\simeq 0$. The third case is the potential $V_0(1-\mu\phi^{-2})$
which comes out of the original mutated hybrid inflation model
\cite{mutated}.
The fourth is the cubic potential advocated in Ref.~\cite{grahamnew}.

With this last potential, for small $N$, 
the scale dependence of the spectrum
(not well-represented by constant $n$) becomes 
strong enough to give a useful lower bound on $N$. It was estimated 
at the end of Section II as $N\gsim 11$.

To summarize the situation regarding $n$ in these models, it
is clear that a measurement of it will
give valuable discrimination between different potentials.
Values of $n$ 
significantly  above
1 are disfavoured theoretically.

We have looked briefly at inflation model-building in the demanding
context of supergravity, focussing on the problem of keeping the 
inflaton mass small enough in the face of generic contributions
of order $\pm H^2$.

We also considered models with a multi-component inflaton, 
and we have looked in some detail at the calculation 
of the spectrum in both this and the single-component case.
In both cases the most powerful calculational technique
\cite{salopek95,ewanmisao} starts with the observation
that after smoothing the 
relevant quantities on scales far outside the 
horizon, the evolution of the universe along each comoving worldline 
will be the same as for an unperturbed universe with the same
initial inflaton field. In the single-component case this justifies
the usual assumption that $\cal R$ is constant. In the multi-component 
case it leads to a simple formula for $n$, whose only input is the 
unperturbed evolution. Finally, we looked at the case of an isocurvature 
perturbation, giving a simple derivation of the previously obscure
fact that the low multipoles of the cmb anisotropy are six times as big 
as for an adiabatic density perturbation.

\begin{table}
\centering
 \caption[table]{Predictions for $n$ are displayed for
inflationary potentials of the 
form $V=V_0(1\pm \mu\phi^p )$, with the first term dominating.
The sixth row represents the limiting case $p\to 0$ from below,
with the minus sign.
In most cases the prediction
is proportional to $1/N$,
where $N$ is the number of $e$-folds of inflation after cosmological 
scales leave the horizon.
Results are given for $N=50$ and $N=25$, corresponds to different 
cosmologies after inflation.
The predictions for $n$ are quoted to only two 
decimal places, because a better observational accuracy
would be very hard to achieve. The symbols {**} mark 
the rows corresponding to five potentials 
that might be favoured on the basis of
theory. The first of them is expected to give $n$ indistinguishable from
1, 
but the last may or may not depending on how the potential is derived.}
\begin{tabular}{cllllc}
 &$V(\phi)/V_0$ & $(n-1)(N/50)$ & \multicolumn{2}{c}{$n$} & \\
&& & $N=50$ & $N=25$ &
\\ \hline 
&$1+\mu\phi$ & \ \ $.00$ & $1.00$ & $1.00$ &\\
 {**}&$1+\mu\phi^2$ & $1.00$\,? &
\multicolumn{2}{c}{$n=1+4\mu$}&{**}\\
&$1+\mu\phi^3$ & \ \ $.08$ & $1.08$ & $1.16$ &\\
&$1+\mu\phi^4$ & \ \ $.06$ & $1.06$ & $1.12$ &\\
&$1+\mu\phi^\infty$ & \ \ $.04$ & $1.04$ & $1.08$ & \\
\hline 
{**}&$1-\mu\phi^{-0}$ & $-.02$ & \ $.98$ & \ $.96$ &{**}\\
 {**}&$1-\mu\phi^{-2}$ & $-.03$ & \ $.97$ &\ $.94$ &{**} \\
&$1-\mu\phi^{-4}$ & $-.03$ & \ $.97$ &\ $.93$ &\\
&$1-\mu\phi^{\pm\infty}$ & $-.04
$ & \ $.96$ &\ $.92$ &\\
&$1-\mu\phi^{4}$ & $-.06$ & \ $.94$ &\ $.88$ & \\
{**}&$1-\mu\phi^{3} $& $-.03$ & \ $.92$ &\ $.84$ & {**}\\
 {**}&$1-\mu\phi^2$ & $1.00$\,??? & \multicolumn{2}{c}{$n=1-4\mu$}& {**}\\
&$1-\mu\phi$ & \ \ $.00$ & $1.00$ & $1.00$ &\\
\end{tabular}
\end{table}

\newpage

\underline{Acknowledgements}:

I am grateful to CfPA and LBL, Berkeley, for the provision of financial 
support and a stimulating working environment when
this work was started.
I am indebted to Ewan Stewart for many helpful
discussions about supergravity, and about multi-component inflaton 
models of inflation. I have
also received valuable input from Andrew Liddle,
Andrei Linde, Hitoshi Murayama and Graham Ross and David Wands.
The work is partially supported by grants from PPARC and
from the European Commission
under the Human Capital and Mobility programme, contract
No.~CHRX-CT94-0423.

\end{document}